\documentclass{aa}  

\usepackage{natbib}
\usepackage{graphicx}
\usepackage{txfonts}
\usepackage{hyperref}
\usepackage{multicol}
\usepackage{subfig}
\usepackage{float}

\begin{document} 

  \title{Galaxy kinematics in the XMMU J2235-2557 cluster field at z$\sim$1.4 
  \thanks{Based  on  observations  with  the  European  Southern  Observatory Very Large Telescope (ESO-VLT), observing run ID 091.B-0778(B).}}

  \author{J. M. P\'erez-Mart\'inez\inst{1}
           \and B. Ziegler\inst{1}
           \and M. Verdugo\inst{1}
           \and A. B\"ohm\inst{2}
           \and M. Tanaka\inst{3}
          }
      
  \institute{Department of Astrophysics, University of Vienna, T\"urkenschanzstr. 17, A-1180 Vienna, Austria. \email{jm.perez@univie.ac.at}
         \and   Institute for Astro- and Particle Physics, University of Innsbruck, Technikerstr. 25/8, 6020 Innsbruck, Austria.
         \and
             National Astronomical Observatory of Japan 2-21-1 Osawa, Mitaka, Tokyo 181-8588, Japan.
             }
  \date{}

  \abstract
   {}
  {The relationship between baryonic and dark components in galaxies varies with the environment and cosmic time. Galaxy scaling relations describe strong trends between important physical properties. A very important quantitative tool in case of spiral galaxies is the Tully–Fisher relation (TFR), which combines the luminosity of the stellar population with the characteristic rotational velocity (V$_{max}$) taken as proxy for the total mass. In order to constrain galaxy evolution in clusters, we need measurements of the kinematic status of cluster galaxies at the starting point of the hierarchical assembly of clusters and the epoch when cosmic star formation peaks.}
  {We took spatially resolved slit FORS2 spectra of 19 cluster galaxies at $z\sim1.4$, and 8 additional field galaxies at $1<z<1.2$ using the ESO Very Large Telescope. The targets were selected from previous spectroscopic and photometric campaigns as [OII] and $H_{\alpha}$ emitters. Our spectroscopy was complemented with HST/ACS imaging in the F775W and F850LP filters, which is mandatory to derive the galaxy structural parameters accurately. We analyzed the ionized gas kinematics by extracting rotation curves from the two-dimensional spectra. Taking into account all geometrical, observational, and instrumental effects, we
used these rotation curves to derive the intrinsic maximum rotation velocity.}
  {V$_{max}$ was robustly determined for six cluster galaxies and three field galaxies. Galaxies with sky contamination or insufficient spatial rotation curve extent were not included in our analysis. We compared our sample to the local B-band TFR and the local velocity-size relation (VSR), finding that cluster galaxies are on average 1.6 magnitudes brighter and a factor 2-3 smaller. We tentatively divided our cluster galaxies by total mass (i.e., V$_{max}$) to investigate a possible mass dependency in the environmental evolution of galaxies. The averaged deviation from the local TFR is $\langle\Delta M_{B}\rangle=-0.7$ for the high-mass subsample (V$_{max}>200$ km/s). This mild evolution may be driven by younger stellar populations (SP) of distant galaxies with respect to their local counterparts, and thus, an increasing luminosity is expected toward higher redshifts. However, the low-mass subsample (V$_{max}<200$ km/s) is made of highly overluminous galaxies that show $\langle\Delta M_{B}\rangle=-2.4$ mag. When we repeated a similar analysis with the stellar mass TFR, we did not find significant offsets in our subsamples with respect to recent results at similar redshift. While the B-band TFR is sensitive to recent episodes of star formation, the stellar
mass TFR tracks the overall evolution of the underlying stellar population. In order to understand the discrepancies between these two incarnations of the TFR, the reported B-band offsets can no longer be explained only by the gradual evolution of stellar populations with lookback time. We suspect that we instead see compact galaxies whose star formation was enhanced during their infall toward the dense regions of the cluster through interactions with the intracluster medium.}
   {}


\maketitle


\section{Introduction}
\label{S:Intro} 

   In the past years, we have achieved a good understanding of galaxy evolution from both
observations and simulations. To mention a few examples, it is well established now 
that cosmic star formation (SF) rises strongly out to redshift 1 \citep{Lilly96} and then turns into a plateau \citep{Bouwens07}. At similar epochs, quasar (AGN) activity of L$_{\ast}$-galaxies 
peaks \citep{Hasinger05}. Deep fields reveal that z = 1 galaxies can already have obtained 
both regular elliptical and spiral morphologies. Hubble volume simulations are able to
reproduce these basic facts \citep{Angulo10}. On the other hand, difficulties persist 
in some aspects like 
the mass-dependent shutting-down of SF by feedback processes \citep{Bower12} and the gas infall 
rate sustaining too high SF rates \citep{Tonini11}. While it is possible to empirically model
certain aspects \citep{Peng10}, we still lack knowledge of the underlying physical mechanisms. 
For example, the relative contributions to galaxy mass growth by major mergers and 
gas accretion \citep{Dekel09} are still controversial.

A very important quantitative tool in case of spiral galaxies is the Tully-Fisher relation (TFR), which 
combines the luminosity of the stellar population (SP) with the characteristic rotational velocity taken as proxy for the total mass (including dark matter). It is well established in the local Universe \citep{Pierce92} and was examined for evolution in recent years out to redshift 1, including work by our own group (\citealt{Ziegler03}, \citealt{Boehm04}, \citealt{Boehm07} and \citealt{Boehm15}). 

The traditional TFR is a subset of a more fundamental relation between baryonic mass (stellar and gas content) and rotational velocity. However, the gas mass fraction can only be observed directly for $z\sim0$ galaxies. For nearby massive $L^{\ast}$-galaxies only a mild brightening was detected that translates into little overall evolution in the baryonic TFR (\citealt{Mcgaugh00}). At higher redshift, \cite{Puech10} constructed this relation by converting multicolor photometry into stellar masses assuming some SP model and estimating the gas mass fraction, finding no evolution up to $z\sim0.6$. This indicates that a significant fraction of spirals have well-established disks at z=1 and do
not undergo major mergers until $z=0$. \citet{Dutton11} furthermore showed that the observed evolution of the scaling relations involving circular velocity, mass, and size are consistent with a simple CDM-based model of disks growing inside evolving Navarro-Frenk-White
(NFW) dark matter haloes. 
However, challenging measurements of 2D velocity fields 
at $z\approx0.5$ reveal disturbances that can be explained by ongoing mass growth through accretion or 
minor mergers (\citealt{Puech08}, \citealt{Kutdemir08}, \citealt{Kutdemir10}). 

In contrast, massive SF galaxies around $z\geqslant2$ can display various kinematic behaviors from disk rotation through dispersion 
dominance to major mergers (e.g., \citealt{Genzel06}, \citealt{Forster09}, \citealt{Lehnert09}). It was recently found by \citet{Wisnioski15} using data from the $KMOS^{3D}$ survey that 93$\%$ of galaxies at $z\sim1$ and 74$\%$ of galaxies at $z\sim2$ are rotationally supported, as determined from a continuous velocity gradient and $v_{rot}/\sigma_{0}>1,$ while they showed a disk fraction of 58$\%$ when applying the additional stricter criteria that the projected velocity dispersion distribution peaks on or near the kinematic center, the velocity gradient is measured along the photometric major axis (for inclined systems), and the kinematic centroid is close to the center of the galaxy continuum. In contrast, applying the morpho-kinematic classification to a $KMOS^{3D}$ subsample, \citet{Rodrigues16} found that only 25$\%$ of
$z\sim1$ galaxies are virialized spirals according to their morpho-kinematic classification. \citet{Simons16} studied 49 galaxies from CANDELS fields with the Keck/MOSFIRE spectrograph and found that high-mass galaxies ($\log{M/M_{\odot}} > 10.2$) at $z\sim2$ are generally rotationally supported and fall on the TF relation, 
while in contrast, a large portion of less massive galaxies at this epoch are in the early phases of assembling their disks.

A sample of 14 galaxies at $z\sim2$ was studied by \citet{Cresci09}, who found a significant offset in the $M_{\ast}$-TFR with increased scatter compared to local samples, which is even more prominent in the z = 3 study of \citet{Gnerucci11}. This implies a strong evolution within 2–3 Gyr with massive SF galaxies changing their dynamical state dramatically before settling into a more quiescent epoch at z=1. 
This demands thorough measurements of spatially resolved emission lines of galaxies 
at z=1–2 in order to understand this transition and its possible causes. A first study by 
\citet{Miller12} exploited ultradeep Keck spectroscopy of the most suitable targets in five deep 
fields (EGS, SSA22, GOODS N\&S, and COSMOS) that benefit from multiwavelength coverage. They 
found that most galaxies follow an $M_{\ast}$-TFR with a mild offset but strongly increased scatter 
compared to the local TFR. In a similar way, \citet{Vergani12} found a lack of any strong evolution 
of the fundamental relations of star-forming galaxies in at least the past 8 Gyr using a sample of 46 galaxies at 1<z<1.6 from MASSIV (Mass Assembly Survey with SINFONI in VVDS). In contrast, \citet{Tiley16} found an offset of the TFR for rotationally supported galaxies at $z\sim1$ to lower stellar mass values (-0.41 dex) for a given dynamical mass but no significant offset in the absolute K-band TFR over the same period, contrary  to  some  previous  studies  conducted at similar redshift but in agreement with the predictions of hydrodynamical simulations of EAGLE. 

Most of these studies were restricted to the field population, while in clusters, additional specific effects should affect the content and structure of galaxies. Galaxy clusters provide special environments to test galaxy evolution across different cosmic epochs. Compared to the field, the number density of galaxies is high, as are the relative velocities. The gravitational potential of a cluster is filled by the intracluster medium (ICM), a hot X-ray emitting gas, and the overall mass-to-light ratio is much higher than for the individual galaxies, indicating the presence of vast amounts of dark matter. This environment exerts a strong influence on the evolution of the cluster galaxies superposed on the (field) evolution that arises from the hierarchical growth of objects and the declining star formation rates over cosmic epochs. In addition to tidal interactions between galaxies, including merging, cluster members are affected by cluster-specific phenomena related to the ICM (such as ram pressure stripping) or the structure of the cluster (such as harassment). Imprints of these interactions can be seen not only in present-day clusters, but they also manifest themselves in a strong evolution of the population of cluster galaxies. One example is the photometric Butcher-Oemler effect of an increasing fraction of blue galaxies with redshift (\citealt{Butcher78}), implying a rising percentage of star-forming galaxies. 

However, comparisons between the TFRs of cluster and field galaxies show no clear results. \cite{Milvang03} and \cite{Bamford05} found higher B-band luminosities in clusters compared to the field, while \cite{Moran07} presented a larger scatter for cluster galaxies. On the other hand, cluster and field populations follow the same trends with no significant differences between these two environments according to \cite{Ziegler03}, \cite{Nakamura06}, \cite{Jaffe11}, \cite{Mocz12}, and \cite{Bosch2}. These discrepancies may stem from the necessity to use only regular rotation curves (RC) for a proper TF analysis that is based on the virial theorem. 

In order to extend such measurements to higher redshifts and to investigate possible biases, we here present a kinematic study of the massive cluster XMMU J2235-2557 (\citealt{Mullis05}) at $z\sim1.4$. 
Making use of the multiwavelength data including HST imaging (\citealt{Rosati09}, \citealt{Strazzullo10}), which allows determining morphologies and accurate deriving of structural parameters (needed for a proper kinematic analysis), we scrutinize the environmental dependence of disk galaxy scaling relations at the highest redshift to date. 

The structure of this paper is as follows: In $\text{Sect.}$ 2 we describe the target selection, observation conditions, and spectroscopic data reduction. The  description of the photometric properties of our sample and details on the derivation of the structural parameters and maximum rotation velocities are shown in $\text{Sect.}$ 3. We present and discuss our results in $\text{Sect.}$ 4 and $\text{Sect.}$ 5, followed by a short summary in $\text{Sect.}$ 6. Throughout this article we assume a \citet{Chabrier03} initial mass function (IMF) and adopt a flat cosmology with $\Omega_{\lambda}$=0.7, $\Omega_{m}$=0.3, and $H_{0}$=70 km s$^{-1}$Mpc$^{-1}$. Magnitudes quoted in this paper are in the AB system.

\section{Sample selection and observations}
                
We carried out multi-object spectroscopy (MXU) with FORS2 between September 2013 and July 2014 to obtain the spectra of 27 galaxies with one mask. We chose the holographic grism 600z+23,
which covers $\sim3300\AA$ in the wavelength range $7370-10700\AA$. The slits were tilted and aligned to the apparent major axis of the targets in order to minimize geometrical distortions. Slit tilt angles $\theta$ were limited to $\left |{\theta}\right |$ < 45º to ensure a robust sky substraction and wavelength calibration. We used a slit width of 0.7", which  delivers an instrumental resolution of $ \sigma_{ins} $= 65 km/s. This configuration yielded a spectral resolution of R $ \sim $ 1400 and an average dispersion of 0.81 \AA/pix with an image scale of 0.25"/pixel.  The total integration time for the MXU observations was 9h per target. In order to diminish the number of cosmic ray hits in our spectra, the observations were divided into 12 observing blocks (OBs) of one hour each, with three subexposures of 15 min per OB plus overheads. We achieved seeing conditions of 0.73 arsec FWHM on average.
   
   The primary targets for the kinematic analysis were 15 cluster galaxies with measured spectroscopic redshifts and [OII]$\lambda$3727 {\AA} emission. These galaxies were extracted from two catalogs of previous spectroscopic campaigns in the same cluster field provided by M. Tanaka and V. Strazzullo (\emph{priv. comm.}). Another set of 5 galaxies with photometric redshift from deep narrow-band imaging corresponding to the rest-frame wavelength of $H\alpha$ were selected from \citet{Grutzbauch} at the cluster redshift. The remaining available mask space was filled with galaxies of disk-like appearance and appropriate position angle, but unknown redshift, yielding 27 targets in total. 
   
   We performed the spectroscopic data reduction mainly using the ESO-REFLEX pipeline for FORS2 (version 1.19.4). The main reduction steps were bias subtraction, flat normalization, and wavelength calibration. The last was improved by removing some lines from the catalog of arc lines. Additionally, we performed bad pixel and cosmic ray cleaning by coadding the exposures with a sigma-clipping algorithm using IRAF. We show the coordinates, redshifts, rest frame colors, and magnitudes of our final sample in Table \ref{tabtot}.

  \section{Methods} 
  \label{SS:Methods}
  
  \subsection{Imaging and photometry}
  \label{SS:Imaging}
  
   To complement the spectroscopy, we made use of imaging data from a variety of sources, including HST/ACS (F775W and F850LP), VLT/FORS2 (B, R, z-bands), VLT/VIMOS (U band), VLT/HAWKI (J and Ks bands), CTIO/ISPI (H band), and Spitzer IRAC (3.6$\mu$m and 4.5$\mu$m), encompassing thus from the rest-frame UV to the near-infrared
(NIR) at the cluster redshift. The characteristics of these datasets are described in Table \ref{T:imaging}.
  
   The HAWKI data reduction is described in  \citet{Lidman08} and the processed images were subsequently released as Phase 3 products in the ESO archive, from where we retrieved them. Zero-points were also provided in the Vega system, which we transformed into the AB system. 
  
   The FORS2 and VIMOS images were also retrieved from the ESO archive as raw data. Similarly, the CTIO-Blanco/ISPI H-band raw frames were downloaded from the NOAO science archive. These datasets were processed with the {\sc Theli} pipeline (\citealt{Schirmer2013}), which takes care of all basic reduction steps as well as the astrometric calibration and coaddition. Photometric calibration for the FORS2 B and R and the VIMOS U-band images was performed using a two step approach. First, zero-points were fixed to the official zero-points available at the ESO webpages. However, noticeable differences were detected in the color of stars in comparison to stellar libraries (\citealt{Pickles1998}). We therefore produced synthetic colors for all available bands and  compared them to the observed colors using the stellar locus method (e.g., \citealt{Kelly2014}), adjusting them until all differences were minimal. The applied corrections ranged from 0.3 to 0.6 magnitudes. 
   
   There was no zero-point available for the FORS2 z band. Fortunately, the filter transmission curve is nearly identical to the HST/ACS F850LP filter available for a significant part of the field. Therefore the FORS2 z band was calibrated against that latter dataset. The ISPI H band was calibrated using 2MASS stars available in the field and transformed to the AB system. The remaining space-based images were retrieved fully processed and calibrated, so that no additional steps were necessary before performing the photometry. 
   
\begin{figure*}
  \centering
  \includegraphics[width=2\columnwidth]{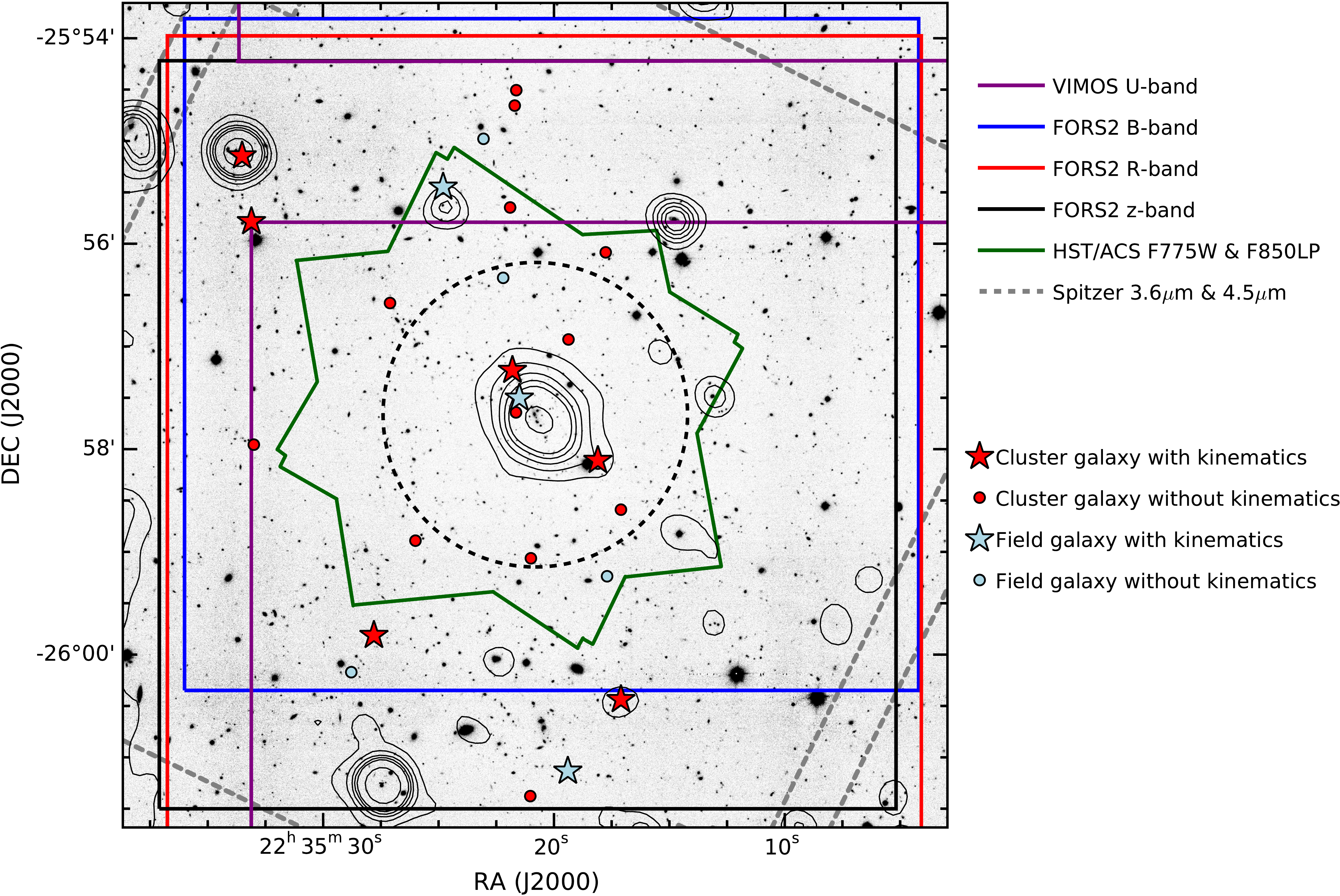}
  \caption{Portion (8.25$\times$8.25\,arcmin$^2$) of the J-band HAWKI image with XMM-$Newton$ X-ray contours overlayed. The field of view of the different instruments used in this work is also shown. The HAWKI J and Ks and ISPI H-band observations cover a field of view much larger than the figure. The FORS2 B-, R- and z-band coverage was obtained from different ESO programs. Most of the X-ray sources are likely distant AGNs, but the extended emission associated with the cluster can be appreciated in the center. The dashed circle marks the $R_{500}=0.75$\,Mpc radius derived by \citet{Rosati09}. We denote the position of the galaxies analyzed in this study by distinguishing between field and cluster and between those that yielded a V$_{max}$ value and those that did not.}
  \label{F:observations}
  \end{figure*}

   The spatial coverage of the different datasets is shown in Fig.\,\ref{F:observations}, where we also mark the targets of the spectroscopic campaign. Clearly, not all galaxies have measurements in all bands. However, they do have enough measurements across a wide wavelength range to reliably determine all parameters necessary for our analysis.
   
  \begin{table*}[!t]
  \caption{Summary of the imaging data used in this work}
  \centering
  \begin{tabular}{llcccc}
  \hline\hline 
   \noalign{\vskip 0.1cm}
  Telescope & Instrument & Filter   & Exp. Time &  PSF FWHM & PID \\ 
            &          &            & (s)       &   (")   &          \\ \hline 
            \noalign{\vskip 0.1cm}
   VLT      & VIMOS    &  U         &  21\,600  &  0.80   &  079.A-0758      \\ 
\ldots      & FORS2    &  B         &   1\,590  &  0.72   &  087.A-0859      \\
\ldots      &  \ldots  &  R         &   2\,250  &  0.75   &  072.A-0706, 073.A-0737   \\ 
\ldots      &  \ldots  &  z         &   1\,200  &  0.40   &  274.A-5024, 091.B-0778  \\  
\ldots      & HAWKI   &  J         &  10\,560   &  0.47   &  060.A-9284(H) \\  
\ldots      &  \ldots  &  Ks        &  10\,740  &  0.32   &  \ldots             \\  
CTIO/Blanco &  ISPI    &  H         &   1\,200  &  1.07   &  2009B-0484        \\ 
 HST        &  ACS/WFC & F775W      & 13\,500   &  0.10    &  10496, 10531, 10698    \\ 
\ldots      & \ldots   & F850LP     & 14\,400   &  0.10    &  \dots             \\ 
Spitzer     &   IRAC   & 3.6$\mu$m  &  6\,262   &  1.95   &  20760            \\ 
\ldots      &   \ldots & 4.5$\mu$m  &  6\,262   &  2.02   &  \ldots          \\ \hline
  
  \end{tabular}
 
  \label{T:imaging}
  \end{table*}
  
Because of the varying depth, field of view, point-spread function
(PSF), pixel scales, and quality of the imaging, special care must be taken in performing the photometry for our target galaxies. 
We therefore deviated slightly from the standard approaches that use more homogeneous datasets. In particular, we chose to forego
degrading the high-quality imaging (HST and HAWKI) to the worst seeing. We did not rebin the images to match the pixel sizes of the rest of the imaging either, as required by dual-band photometry. Instead, we chose to measure magnitudes within an elliptical Kron aperture (\citealt{Kron1980}) matched to the seeing in each image. To calculate the size of the Kron aperture, we used {\sc SExtractor} (\citealt{Bertin1996}) in the high-quality data where we measured the coordinates, the Kron radius, the ellipticity, the position angle, and the Kron magnitudes ({\tt MAG\_AUTO}).  These parameters where passed to the python {\sc phot\_utils} tools, which replicates many of the functionalities found in {\sc SExtractor} in a more flexible environment.  

To calculate by how much the Kron apertures need to grow to take into account the different seeing values in the other bands, we used the software {\sc Stuff} and {\sc Skymaker} (\citealt{Bertin2009}) to simulate realistic galaxy fields with different spatial resolutions, where we ran  {\sc SExtractor}  with the same parameters over the same galaxies. In general, the Kron radii growth can be modeled as a simple linear function that depends only on the measurements in the high-quality imaging and the seeing in the lower quality imaging. 
Magnitudes computed by {\sc phot\_utils} using the derived apertures are in excellent agreement (at $\sim$0.1\,mag level) with those determined directly by {\sc SExtractor}. 

The above procedures were not applied to the Spitzer images,
however. With a PSF FWHM of $\sim$2\,arcsec, distant galaxies are effectively unresolved in the IRAC images. We therefore used a fixed circular aperture of 6\,arcsec and applied the standard correction factors for the missing flux  (factors 1.205 and 1.221 in 
the 3.6 and 4.5\,$\mu$m bands, respectively). 
In conclusion, we estimate the total calibration for all bands to have an accuracy of 0.1 magnitudes.

\subsection{Stellar masses and rest-frame magnitudes}
  
Rest frame magnitudes and stellar masses were determined using the code {\sc Lephare} of  \citet{Arnouts2011} 
(see also \citealt{Ilbert2006}), which fits stellar population synthesis models (\citealt{BC03}) to the
spectral energy distribution (SED) derived from the photometry. The code is a simple $\chi^2$ minimization algorithm that finds the best match of templates for the given data. To avoid overfitting, we restricted the possible ages to values lower than the age of the Universe at the redshift of the cluster. 
Thus, we have average errors in absolute magnitude and stellar masses of 0.12 mag. and 0.09 dex, respectively.

\begin{figure}
   \centering
   \includegraphics[width=\hsize]{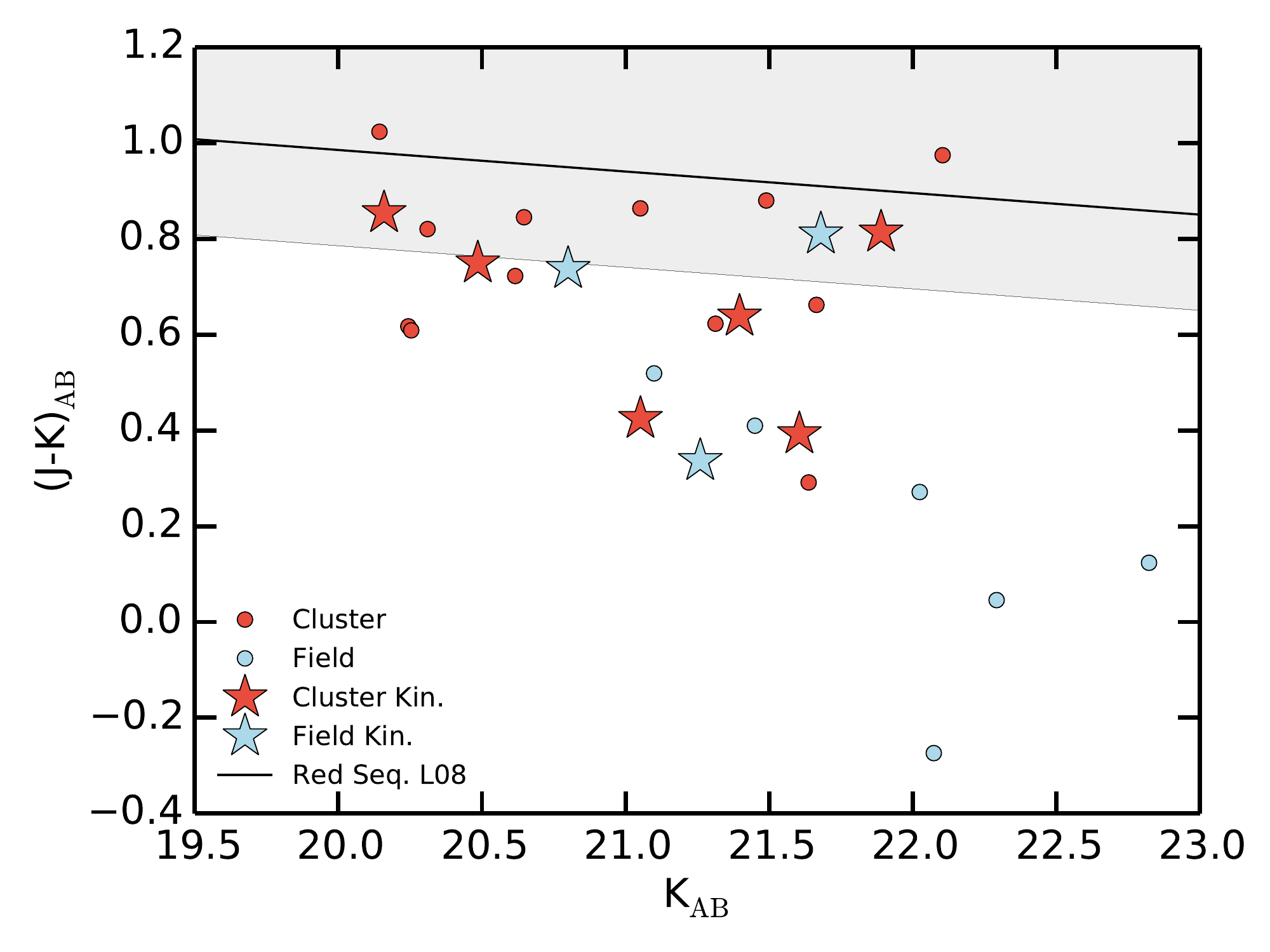}
      \caption{Color-magnitude diagram. Red stars and circles define spectroscopic cluster members with and without derived kinematics in our sample, respectively. Blue stars and circles show field galaxies with and without derived kinematics in our sample, respectively. 
      The cluster red-sequence fit derived by \citet{Lidman08} is shown as a black line with a shaded area: red-sequence galaxies are defined as galaxies redder than 0.2 magnitudes blueward of this fit, which is shown with a shaded area. Magnitudes and colors are given in the AB system.}
         \label{magcolor}
   \end{figure}

  To place our sample of cluster galaxies into context, we present in Fig. \ref{magcolor} the color-magnitude diagram in J and K
bands for XMMJ2235 galaxy members. The cluster red-sequence fit from \citet{Lidman08} is shown with a black line, with red-sequence galaxies defined as galaxies redder than 0.2 magnitudes blueward of this fit. The purpose of this comparison is to highlight the nature of our cluster galaxies with [OII]-based kinematics. 
  We would like to mention that two of our galaxies (IDs 8 and 11 in Table \ref{tabtot}) were part of the Herschel sample of dust-obscured star-forming galaxies presented in \citet{Santos13}. 
  
  Edge-on disks have higher extinction (A$_{B}$) than face-on galaxies, and more massive disks are dustier than less massive disks \citet{Giovanelli95}. We corrected the rest frame B-band absolute magnitudes for intrinsic dust absorption following the approach by \citet{Tully98}:  
 
  \begin{equation}
  A_{B}=\log (a/b) [-4.48+2.75\log(V_{max})]
  .\end{equation}
  
  The extinction is dependent on the inclination angle $i$, which is related to the ratio between the axes $(a/b)$, and on the $V_{max}$ of every galaxy. After applying the extinction correction, the typical errors in B-band luminosity range from 0.2 to 0.3 mag.

  \subsection{Structural parameters}
  
  Owing to the relatively small spatial coverage of the HST images in this cluster, we were able to derive the structural parameters of galaxies from the z-band (F850LP) images only in the central regions of the cluster. For galaxies far from the cluster center, we used ground-based HAWKI photometry in K band. We used the GALFIT package by \citet{Peng02}, which allows fitting multiple 2D surface brightness profiles simultaneously to the galaxy under scrutiny. We fit the surface brightness profiles of the galaxies of
  our sample using two different methods: 
  
   \begin{enumerate}
      \item A single Sérsic profile with free index $n_{s}$.
      \item A two-component model with an exponential profile ($n_{s}=1)$) for the disk and a de Vancouleurs profile ($n_{s}=4$) for the bulge.
   \end{enumerate}

     The best parameters from the first method were used as initial guess values for the second. All fit residuals were visually inspected (Fig. \ref{foot}), and in a few cases, it was  necessary to apply constraints on some parameters in order to avoid a local minimum in the fitting process. We used the bulge/disk decomposition to obtain the disk parameters when possible within the HST coverage. However, an accurate structural decomposition was not feasible with the ground-based K-band data because of the limited spatial resolution. We therefore restricted the GALFIT models to a single Sersic profile for galaxies without available HST imaging. We stress that for the analysis presented here, the most important parameters are the ratio between the axes $(a/b)$, position angle $\theta$, and scale length $R_{d}$ of the disk.

  However, observed scale lengths depend on the wavelength regime. It was shown in \citet{Jong96} that the scale length ($R_{d}$) is smaller when galaxies are observed in redder filters. The HAWK-I K-band photometry  overlaps with the HST z band over a region where a subsample of 14 galaxies can be studied. We carried out the structural parameter determination in both bands, finding that HAWKI K-band-based effective radii ($R_{e}$) are on average $25\%$ smaller than z-band-based radii. We corrected HAWKI based $R_{e}$ onto z band to make them comparable before computing the scale lengths. 
  GALFIT only returns random errors on the best-fit parameters. These are very small (<1 $\%$) throughout our sample. We relied on a previous analysis of HST/ACS images using GALFIT in \citet{Boehm13} to obtain a more realistic estimate of the systematic errors on $R_{d}$. In that work, a typical systematic error of $20\%$ on galaxy sizes was found for a negligible central point source. This value hence represents the systematic size error for galaxies with the light profiles of pure disks or disks with only weak bulges; this is the case for the vast majority of galaxies  in  our  sample.  We  therefore  adopt a $20\%$ error on $R_{d}$ in the following.

  The inclination $i$ is the angle between the normal vector of
  the disk and the line of sight. We computed it from the apparent major axis a and the apparent minor axis b following \citet{Heidmann72}:
  \begin{equation}
  \cos^2{(i)}=\frac{(a/b)^2-q^2}{1-q^2}
  .\end{equation}
  Here the factor q is the ratio between disk scale length and scale height and is fixed to 0.2, which is the 
  observed value for typical spirals in the local Universe \citep{Tully98}. However, at high redshift, the intrinsic disk thickness (parameter $q$) may be different. Since GALFIT fitting parameter errors are negligible, the uncertainty in the assumption of parameter q is the main source of error in the determination of $i$. We allowed different  values for the disk thickness in order to simulate a thick disk ($q=0.3$) and an infinitely thin disk ($q=0$). The systematic error on the inclination due to the different triaxial configurations is of a few degrees ($<5º$) for high-$i$ galaxies and almost negligible for low-$i$ galaxies.
  
  The position angle $\theta$ denotes the orientation of the apparent major axis in the plane of the sky. Throughout 
  this paper, the convention is that $\theta$ gives the angle between the apparent major axis and the horizontal 
  axis, counted counter-clockwise. To minimize the geometric distortions, we constrain the mismatch angle $\delta,$ which gives the deviations between the major axis and the slit direction of a certain galaxy to $\delta \leq 30^{\circ}$.
  
  \subsection{Rotation-curve extraction and modeling}
   
   Our approach to extract rotation curves from spatially resolved spectra and determine V$_{max}$ is explained in detail in \citet{Boehm04}, \citet{Bosch2}, and \citet{Boehm15}. A summary of the main steps is presented here.
   
   Before the emission line fitting, we use an averaging boxcar of three pixels width, corresponding to 0.75", for
   each spatial position in the spectrum to enhance the signal-to-noise ratio (S/N). We then transform red- and
   blueshifts of the emission lines as a function of galactocentric radius into an observed position-velocity diagram. The kinematic center is determined by minimizing the asymmetry of the observed RC, with a maximum allowed mismatch between kinematic and photometric center of $\pm$1 pixel, corresponding to $\sim$2kpc at the redshifts of our targets.
  
   To determine V$_{max}$ for a given galaxy, we simulate its rotation velocity field by taking into account all geometric effects such as disk inclination angle and observational effects like seeing or the influence of the slit width. The simulated velocity field is generated by assuming a linear
   rise of the rotation velocity $V_{rot}(r)$ at $r<r_{t}$, where $r_{t}$ is the turnover radius, and a convergence of
   $V_{rot}(r)$ into a constant value $V_{max}$ at $r>r_{t}$ \citep{Courteau}.
   
   In most cases, the turnover radius was set as equal to the scale length, $R_{d}$, measured from the stellar morphology. However, some galaxies required $R_{d}$ fitting because the stellar scale length and the turnover radius of our extracted rotation curves were mismatched. In the last step we extract from the synthetic
   velocity field a simulated rotation curve from which we obtain the intrinsic maximum rotation velocity $V_{max}$ taking into account the structural and observational parameters. The error budget on $V_{max}$ was computed following Eq. 3 in \citet{Boehm04}, taking into account the error from the $\chi^2$-fits of the synthetic to the observed rotation curve, and the propagated errors of the inclination and the misalignment angle. The typical error on $V_{max}$ is 20-30 km/s. For a complete description of the full process, we refer again to \citet{Boehm04}. The synthetic velocity fields and simulated and observed rotation curves of our sample are shown at the end of this paper.

   \begin{figure*}
    \begin{multicols}{2}
      \includegraphics[width=\linewidth]{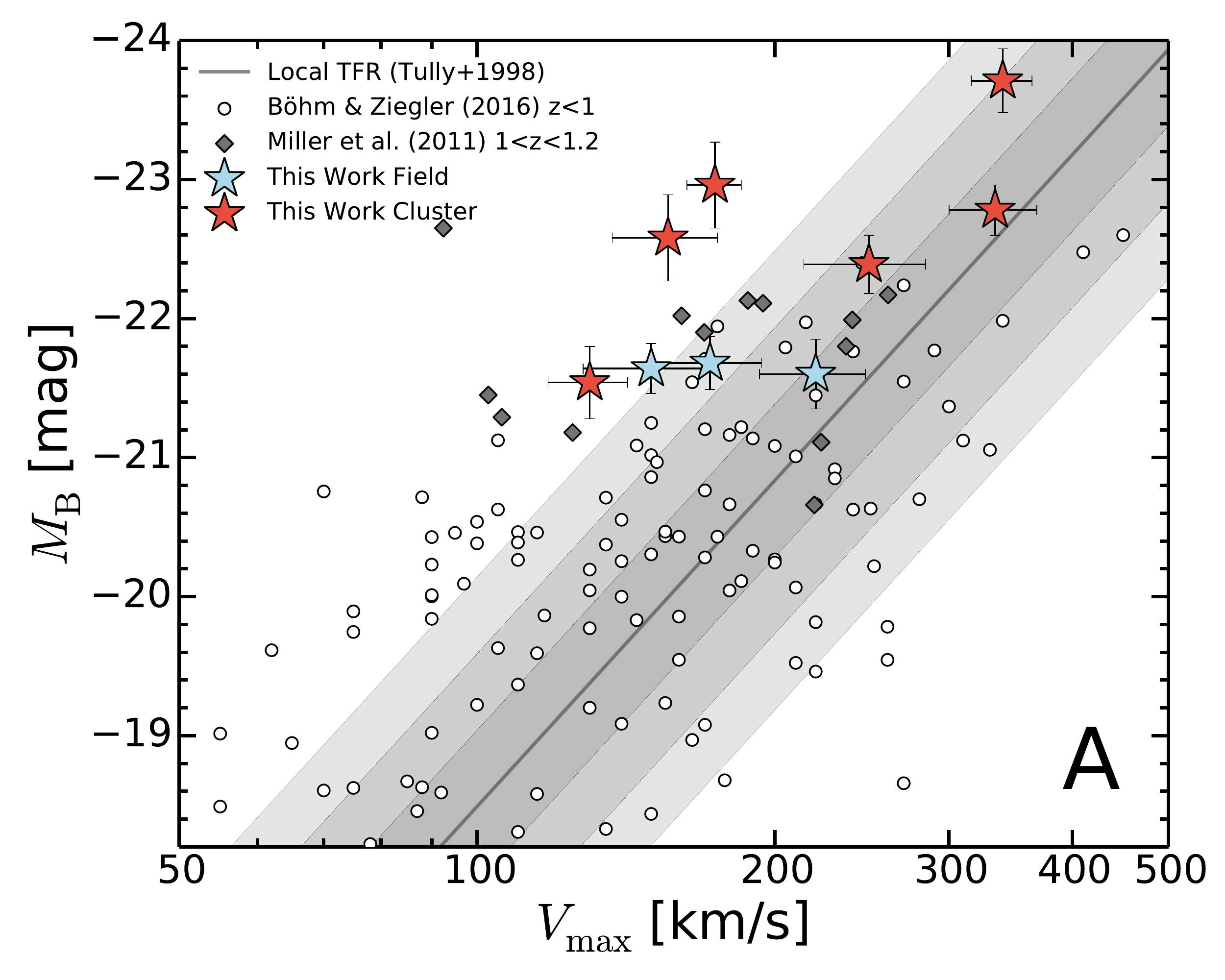}\par 
      \includegraphics[width=\linewidth]{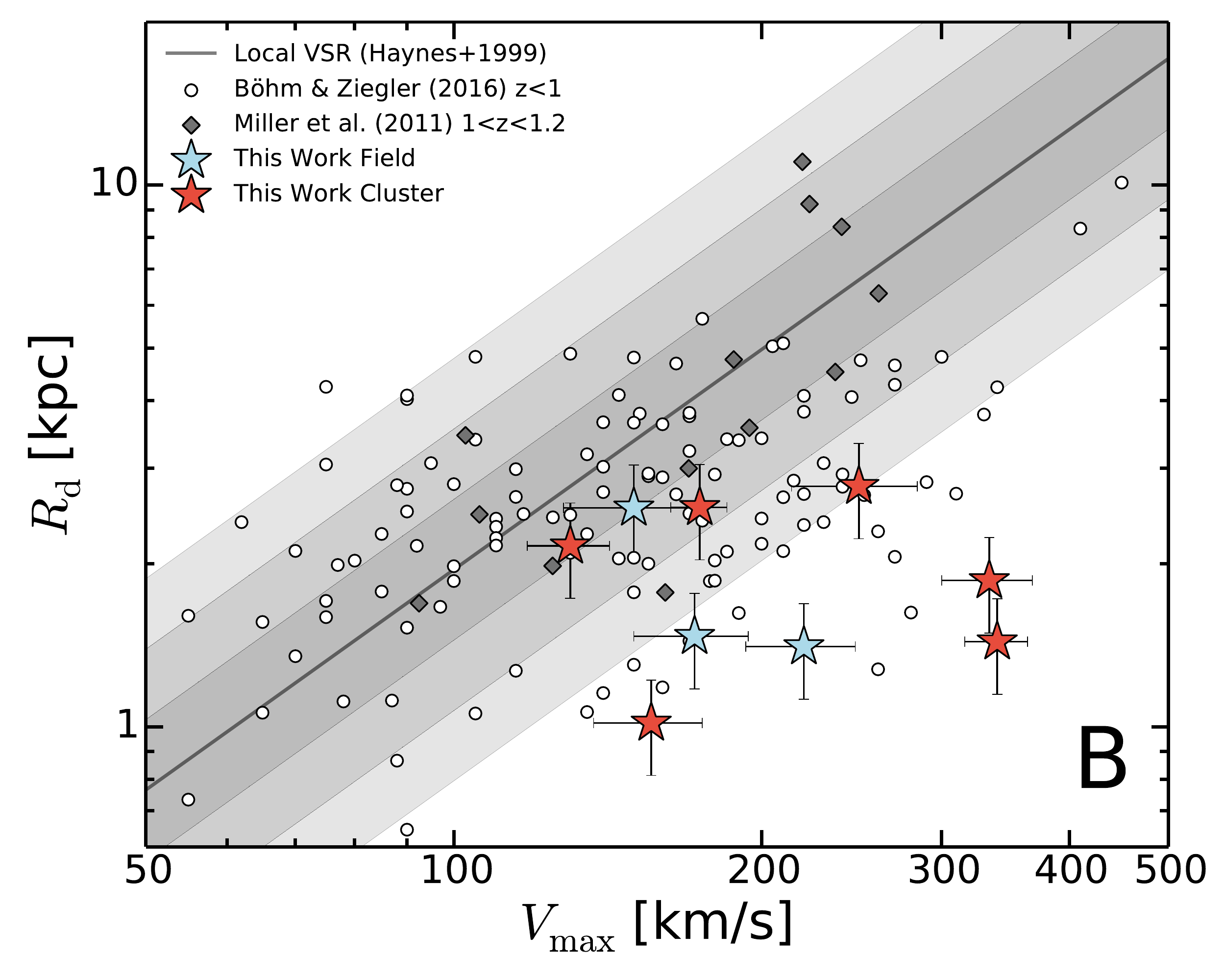}\par 
      \end{multicols}
    \begin{multicols}{2}
      \includegraphics[width=\linewidth]{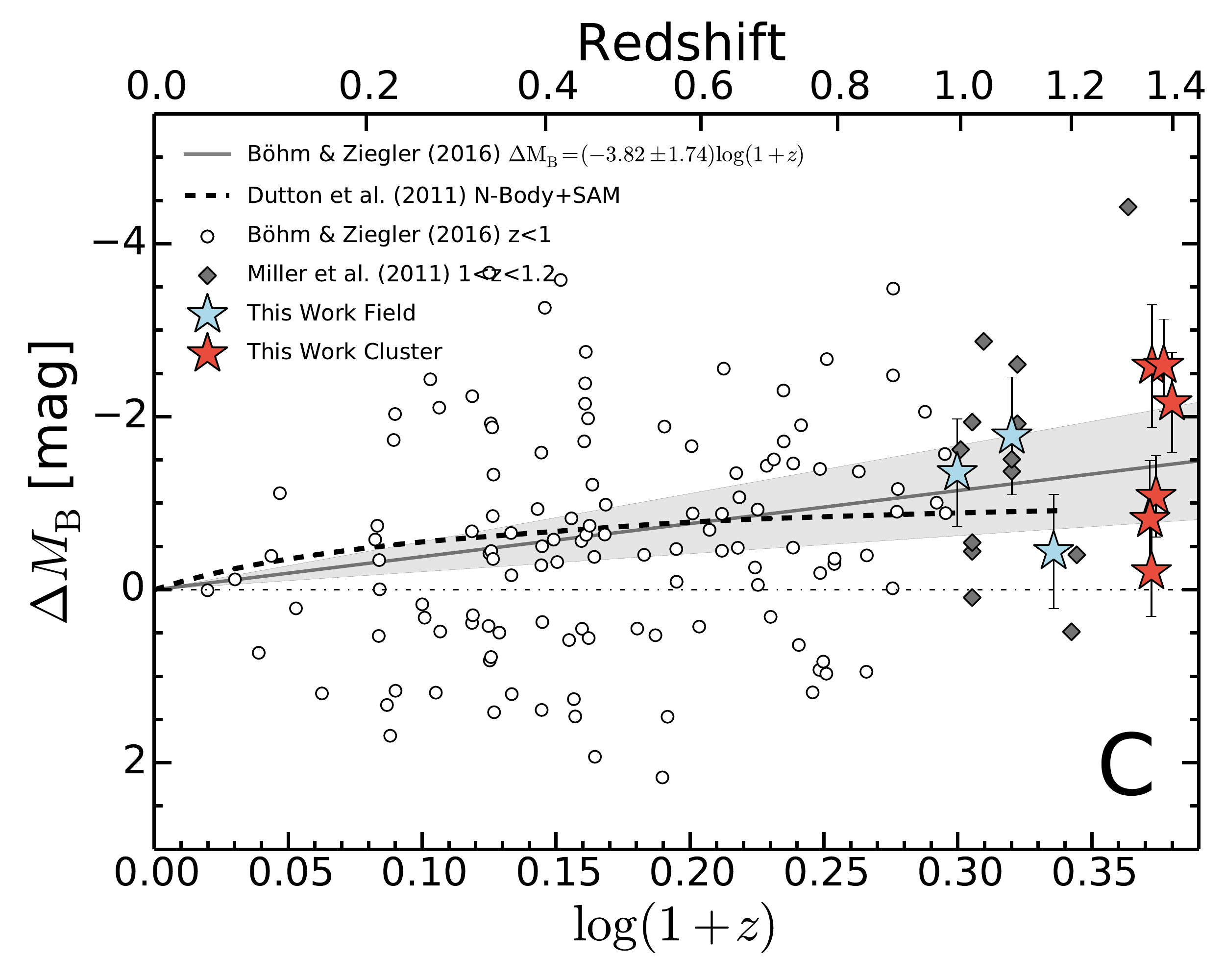}\par 
      \includegraphics[width=\linewidth]{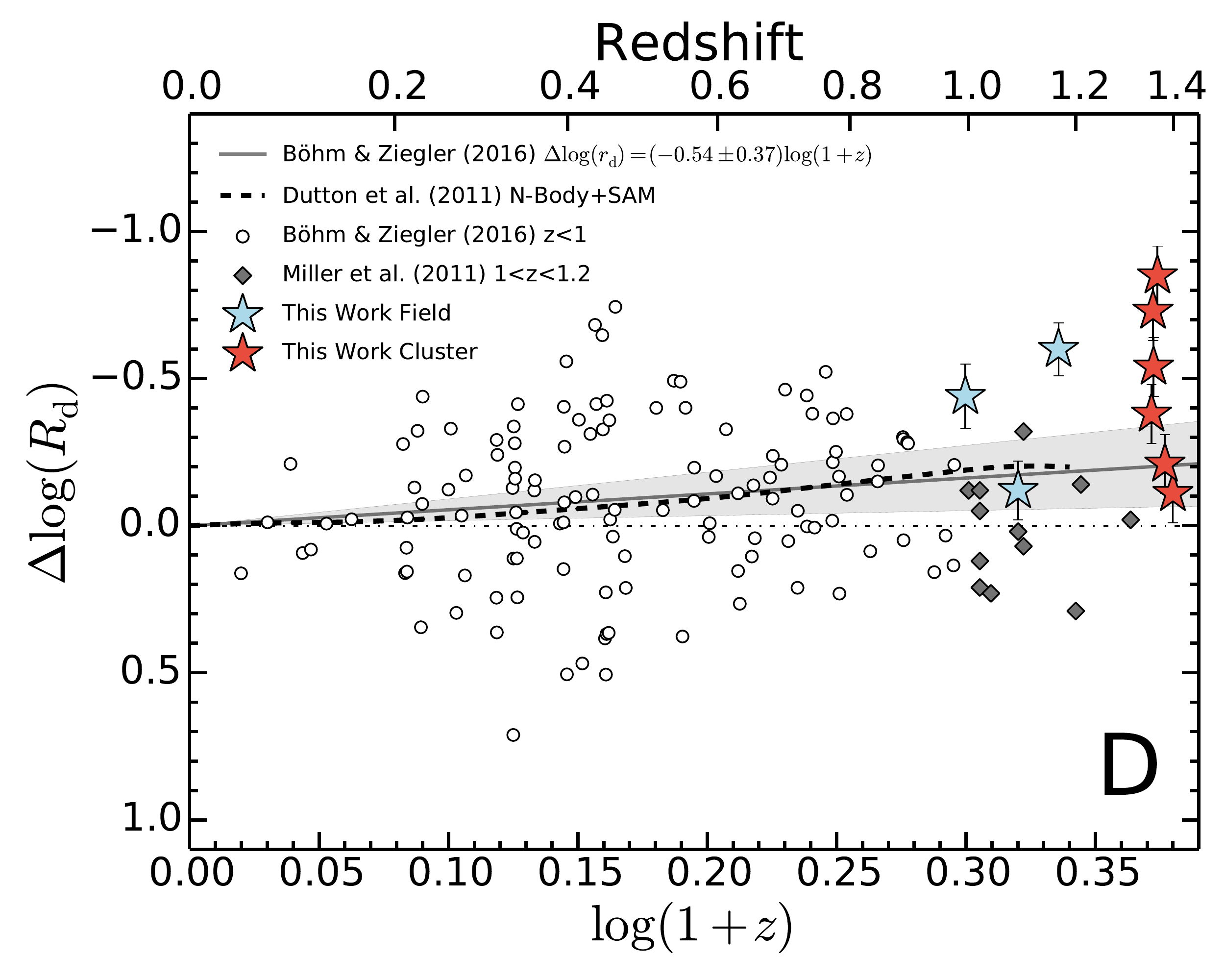}\par 
      \end{multicols}
      \caption{A) Tully-Fisher B-band diagram. B) Velocity-size relation. C) TFR evolution diagram. Offsets $\Delta M_{B}$ of galaxies in our sample from the local TF relation, displayed as a function of redshift. D) VSR evolution diagram. Offsets $\Delta R_{d}$ of galaxies in our sample with respect to the local VSR, displayed as a function of redshift. In A and B the gray line represents the local TFR \citep{Tully98} and the local VSR \citep{Haynes99b}, respectively, with
      1$\sigma$, 2$\sigma,$ and 3$\sigma$ shaded regions. Two samples of field galaxies are plotted for
      comparison: the first comprises 124 disk galaxies out to z=1 from \citet{Boehm15} and is
      plotted with black circles. \citet{Miller11} provided a second field sample composed of 13 disk
      galaxies at 1$<$z$<$1.3 plotted as olive diamonds. The three field galaxies at 1$<$z$<$1.2
      targeted in our observations are plotted with blue stars. Cluster galaxies are represented by red stars.}
         \label{TF_VSR}
      \end{figure*}
   
    Although we observed 27 objects, only 25 of them turned out to be emission line galaxies after the analysis of their spectra. Redshift values were computed using IRAF under visual determination of [OII] emission line center, that is, the only emission line available for cluster members. We detected 17 cluster members and 8 field galaxies. However, part of the cluster sample was affected by strong sky line contamination. As a result, 8 cluster galaxies were discarded because of noisy contamination. We extracted 6 RCs of cluster members from the remaining 9 galaxies, discarding 3 galaxies because of their very compact [OII] emission, which prevents us from reaching the flat part in the RC. 
    In addition, 3 RCs were extracted from the field sample, discarding in the process 3 galaxies with sky contamination and 2 galaxies with compact emission.

\section{Results}
   The primary results of this paper are shown in Fig. \ref{TF_VSR}, where we present the 
   TFR and VSR for cluster galaxies (red 
   stars) at z$\sim$1.4.  
   In Fig. \ref{TF_VSR}A we show the B-band TFR, which is related with recent episodes of star formation. The distribution of our kinematic sample is compared with the local TFR by \citet{Tully98} and the local VSR derived by \citet{Boehm15} using data from \citet{Haynes99a}. In addition, two field samples within 0.2<z<1.3 are shown as comparison between different environments. 
   
   The first field sample comprises 124 disk galaxies out to z=1 from \citet{Boehm15} (hereafter BZ16) plotted as black circles. This is one of the largest kinematic samples of distant galaxies to date. A second field sample composed of 13 disk galaxies at 1<z<1.3 selected by redshift from a larger sample provided by \citet{Miller11} (hereafter M11) is plotted as gray diamonds. We also display the 3 field galaxies targeted in our observations at 1<z<1.2 (blue stars). On average, cluster galaxies in our sample are brighter by $\langle \Delta M_{B}\rangle=-1.6$ mag in B band than in the local TFR and smaller by a factor of 3 than in the local VSR (Fig. \ref{TF_VSR} A and B, respectively). In order to determine environmental effects, we compared the results of our cluster sample with Miller's field galaxies at similar redshift and found that our cluster galaxies are brighter on average by $\langle \Delta M_{B}\rangle=-0.3$ mag in B band and smaller by a factor of 2 than their field counterparts. 
   
   In order to explore a possible environmental mass-dependent evolution and considering the position of our cluster galaxies in the Tully-Fisher diagram (Fig. \ref{TF_VSR}A), our cluster sample was divided into two groups: the first is composed of three high total mass (V$_{max}>200$ km/s) galaxies lying below the $2\sigma$ region of the local TFR with $\langle \Delta M_{B}\rangle=-0.7\pm0.6$. The second group is composed of three low-mass galaxies lying between the 4 and 5$\sigma$ region with respect to the local TFR and showing $\langle \Delta M_{B}\rangle=-2.4\pm0.6$. The cause of this discrepancy between groups might be different physical processes acting on them due to the cluster environment. The errors on $\Delta M_{B}$ are computed through error propagation from the errors on V$_{max}$ and M$_{B}$ according to Eq. 3 in \citet{Boehm15}. Because of the limited size of our subsamples, the uncertainty on $\langle \Delta M_{B}\rangle$ represents the average measurement error.
   
   Figure \ref{TF_VSR}C and D display the offsets $\Delta M_{B}$ and $\Delta \log{R_{d}}$ from the local TFR and VSR as a function of redshift. In panel C field galaxies show increasing overluminosities toward higher redshifts despite the scatter of the samples. This evolution is explained by the rise of SFR and the gradual evolution of SP with lookback time. A simple extrapolation of the luminosity evolution linear fit given by \citet{Boehm15} (gray line) shows that $\Delta M_{B}=-1.2\pm0.4$ magnitudes at z=1.4. As explained in Figure \ref{TF_VSR}A, our cluster sample can be divided into two groups by their total mass. We compare our results with the theoretical predictions of \citet{Dutton11}  (dashed line), who found $\Delta M_{B}=-0.95$ mag at this redshift. Our high total mass group of galaxies is in agreement with the theoretical predictions within the errors, while the low total mass group of galaxies differ by 1.4 magnitudes. In addition, our field sample of galaxies lie within the 1$\sigma$ scatter area of the predicted luminosity evolution shown in \citet{Boehm15}, and at the same time, they are compatible with predictions from numerical simulations by \citet{Dutton11}. 

   In panel D previous samples of field galaxies show decreasing sizes toward higher redshifts, although the scatter of the samples is similar to the scatter found in C. According to the extrapolation of the linear fit given in BZ16 (gray line), the size evolution reaches $\Delta \log{R_{d}}=-(0.22\pm0.14)$ at z=1.4, where negative values in $\Delta \log{R_{d}}$ mean smaller sizes at a given maximum rotation velocity V$_{max}$. In this plot the z=1.4 cluster galaxies do not populate two separate groups, but cover a broad range in size evolution. On average, they are two to three times smaller than their local counterparts, showing $\langle\Delta \log{R_{d}}\rangle=(-0.47\pm0.15)$. Our three field galaxies cover a similar range of scale lengths with a slightly higher mean value, $\langle\Delta \log{R_{d}}\rangle=(-0.39\pm0.14)$. In contrast, predictions from numerical simulations by \citet{Dutton11} (dashed line) showed that $\Delta \log{R_{d}}=-0.2$ at the same redshift.  
   Errors on $\Delta \log{R_{d}}$ are computed through error propagation from the errors on V$_{max}$ and R$_{d}$ following Eq. 6 in \citet{Boehm15}. 

\begin{figure*}
   \centering
   \includegraphics[width=12cm]{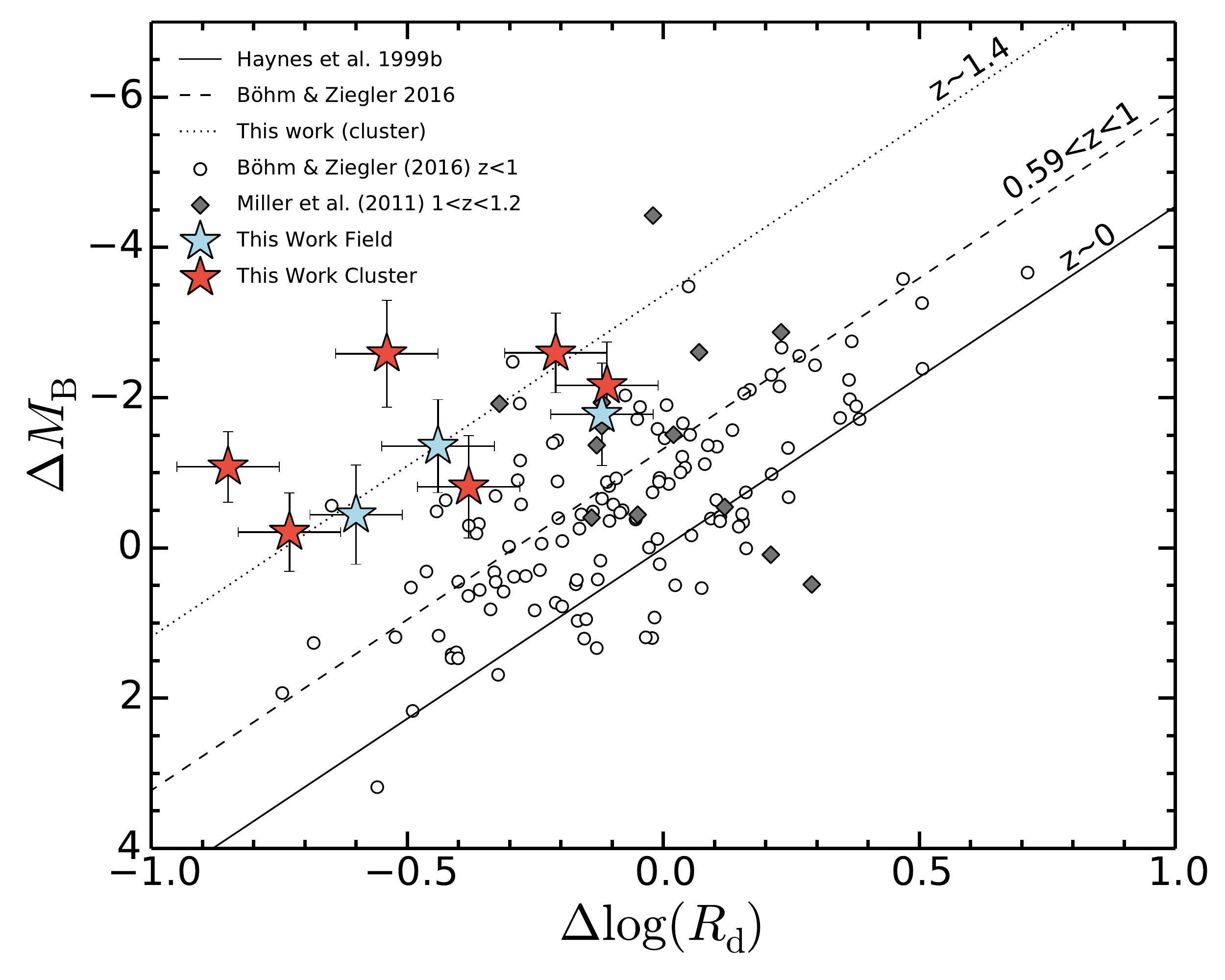}
      \caption{Correlation diagram between the offsets, $\Delta M_{B}$, from the TFR and the offsets, $\Delta\log{R_{d}}$, from the VSR for previously presented samples. The solid line shows the linear fit for galaxies studied in the local Universe \citep{Haynes99a}. By definition, this line goes through the coordinates origin. The dashed black lines plotted in this diagram show the linear fit for galaxies studied in \citet{Boehm15} at 0.59<z<1, showing that galaxies are shifting away with lookback time toward higher luminosity and smaller sizes. The dotted black line is the linear fit at a fixed slope for our cluster sample, whose zero-point is shifted by -0.69 dex in $\Delta\log{R_{d}}$ with respect to the local sample.}
         \label{delta}
   \end{figure*}

   To explore the connection between scaling relations for disk galaxies, we compare in Fig. \ref{delta} the offsets $\Delta M_{B}$ from the TFR with the offsets $\Delta \log{R_{d}}$ from the VSR. By definition, the median of the two parameters is zero in the local Universe. However, there is a clear correlation between $\Delta M_{B}$ and $\Delta \log{R_{d}}$ because luminosity, size, and maximum velocity conform a 3D parameter space in which disk galaxies populate a plane. Thus, these three parameters are correlated, and deviations between local and distant galaxies reflect the evolution of one or several of the parameters. Figure \ref{delta} shows the projection of this 3D space on a luminosity-size plane represented by the offsets in disk scale length and absolute magnitude from the local VSR and TFR. This representation was recently used by BZ16 to quantitatively study the galaxy evolution in the field up to z=1. Using fixed-slope fits to determine the offsets from the local relation in terms of $\Delta\log{R_{d}}$, these authors found a combined evolution in size and luminosity in their sample with a zero-point $\Delta\log{R_{d}}=-0.29$ for field galaxies at 0.59<z<1. 
   Our sample of cluster and field galaxies at 1.2<z<1.4 shows larger offsets and a zero-point of $\Delta \log{R_{d}}=-0.74$. Although our cluster sample has large scatter, our data follow the general trend in the BZ16 sample at 0<z<1: distant galaxies are shifting away from the local $\Delta M_{B}-\Delta \log{R_{d}}$ relation toward smaller sizes and higher luminosity with lookback time. 
   
   \section{Discussion}
  
   The brightening detected in the B-band TFR for cluster galaxies can be partially explained by the increasing star formation galaxies experience toward longer lookback times. However, the division of our sample into two groups according to their total mass and B-band luminosity offsets might indicate additional effects that may explain their properties. Several possible explanations have to be considered.
   
   \subsection{Observational Effects}
  
   First, V$_{max}$ might be underestimated. \citet{Persic96} studied the relation between the mass of a galaxy and the shape of its RC by introducing a complex definition of a universal rotation curve (URC). According to this study, very low-mass  spirals  show  an  increasing rotation velocity even at large radii, whereas the rotation curves of very high-mass spirals moderately decline in that regime. These gradients are found as far as 5 optical disk scale lengths. However, our cluster sample mainly covers intermediate masses ($10.03 <\log{M_{\ast}}< 10.91$), where the URC does not introduce a velocity gradient at large galactocentric radii. 
   However, the spatial extent of the rotation curves in our sample (as well as other samples at similar redshifts) is around two to four times R$_{d}$, which is insufficient to constrain potential RC gradients in the outer disk. With our RCs, we probe out to radii similar to R$_{opt}$, and for galaxies with $V_{rot}(R=R_{opt})>100$ km/s (all of our galaxies), the largest possible underestimate in V$_{max}$ due to RC gradients is $10-20\%$ (see Fig. 4 in \citealt{Persic96}).
   
   We also checked whether these distributions are caused by a selection effect that is due to a magnitude limit in the spectroscopic and photometric campaigns from which we extracted our targets. Toward higher redshifts, such an apparent magnitude limit corresponds to higher luminosities and in turn higher masses. A fraction of the low-luminosity low-mass (slowly rotating) spiral population is therefore missed in the selection process, while the low-mass galaxies that are selected might preferentially be located at the high-luminosity side of the TF relation. This effect is commonly know as Malmquist bias. Thus, in all redshifts bins, at a given V$_{max}$ any distant galaxy sample with a magnitude limit will show an overluminosity of the low-mass galaxies compared to the local TFR, while the distributions are similar at the high-mass end. Some of the previous studies we used for target selection (\citealt{Strazzullo10}, \citealt{Grutzbauch}) have magnitude limits of $z_{AB}=24$ and $H_{AB}=24.4$. However, our sample is well inside the limits showing average observed magnitudes of $z_{AB}=22.4$ and $H_{AB}=21.3$, meaning that the magnitude bias in our target selection should not have a great impact. In addition, we studied the distribution of our targets in B-band luminosity. For the full sample, the mean B-band luminosity value before applying the absorption correction is $\langle M_{B}\rangle=-21.8$ mag with a scatter of $\sigma_{total}$=0.9 mag. As stated in Sect. 2.4, 25 out of 27 galaxies showed [OII] emission, but we did not extract RC from all of them for diverse reasons (OH contamination, compactness, faintness). Neglecting galaxies whose emission lines are contaminated leaves us with a clean sample of 17 galaxies with the same $\langle$M$_{B}\rangle$ and scatter. This means that the removal of OH-affected galaxies does not introduce a luminosity bias to the clean sample. 
   Now, if we focus on the cluster members, we see that they have similar M$_{B}$ , but cover a wide range in V$_{max}$. The high-mass (fast rotating) group shows higher M$_{B}$ than the clean sample, $\langle M_{B}\rangle=-23.0$ with scatter $\sigma_{high}=0.6$, and a high average maximum rotation velocity, $\langle V_{max}\rangle=308$ km/s. On the other hand, the overluminous low-mass (slowly rotating) group presents similar M$_{B}$ , but relatively low V$_{max}$, $\langle M_{B}\rangle=-22.4$ with $\sigma_{low}=0.7$ and $\langle V_{max}\rangle=154$ km/s. Thus, both groups have similar B-band luminosities, but only the low-mass group is significantly offset with respect to the local TFR, 
   which might point toward the presence of a magnitude bias. Nevertheless, the small number of galaxies make it hard to draw firm conclusions, and the large B-band offsets in the TFR ($\langle\Delta M_{B}\rangle=-2.4$ mag for the low-mass cluster galaxies) probably require additional cluster-specific effects to explain the enhanced luminosity.

   \subsection{Physical effects}
   
   The third scenario suggests that we might see rather compact galaxies that became enhanced in SF during their infall toward the dense regions of the cluster. If this is the case, this enhancement should be caused by a process that does not strongly affect galaxy gas kinematics, at least within the galactocentric radial regime probed by our RCs, up to $\sim$3 scale lengths, and during the infall phase where the ICM density has low to intermediate values. 
   
   \citet{Kronberger08b} and \citet{Kapferer09} investigated the influence of ram pressure stripping (RPS) on the internal gas kinematics of simulated spiral galaxies by focusing on how the resulting distortions of the gaseous disk translate into the RC and the full 2D velocity field (VF) of galaxies. Distortions and declining RCs were found at distances larger than 12 kpc from the center of the galaxy, indicating the presence of an undisturbed inner part below that radius. The inclination of the galaxy relative to the line of sight changes the degree of disturbance and may shift the luminosity center from the kinematic center. In our sample the RCs cover radii up to $\sim$10 kpc, and therefore we are not able to investigate possible distortions at larger radii. The absence of irregularities in the inner parts may be a hint toward this type of effect. In addition, the compression of the gas in the central parts that is due to the pressure of the intracluster medium (ICM) can trigger new star formation and a subsequent brightening of the ram-pressure affected galaxy, causing high-luminosity TF offsets. According to \citet{Kronberger08a}, RPS enhances the SFR by up to a factor of 3 over several hundred Myr for a Milky Way-like galaxy. In total, the mass of newly formed stars is about twice higher than in an isolated galaxy after 500 Myr of high ram pressure acting. However, these studies do not reach the ICM density conditions usually found in massive clusters. From the observational point of view, \citealt{Ebeling14} showed that cluster galaxies suffering strong RPS can increase their SF and become temporarily brighter than even the BCG of the cluster. Although this is only expected to occur rarely and only in very massive clusters and for small angles between the normal vector of the disk and the vector of movement through the ICM, several such cases have been discovered (e.g., \citealt{Owen06}, \citealt{Cortese07}, \citealt{Owers12}, \citealt{Ebeling14}). 
   
   However, individual events of this intensity may be rare, requiring a gas-rich galaxy to cross deep within the cluster core at very high velocity. \citet{Ruggiero17} closed this gap by simulating Milky Way-like infalling galaxies in clusters around $10^{14}-10^{15}M_{\odot}$ and choosing $R_{200}$ at present time as the initial density conditions for the ICM at the beginning of the galaxy infall. This study takes into account three different orientations of the galaxy disk (0º, 45º, and 90º) for a radial infall speed of 0.5 to 2 times the velocity dispersion of the cluster. Their results show that star formation is always initially enhanced by a factor of 1.5 to 3 by the compression of the gaseous disk. Interestingly, the SFR  increases by a factor of 2 before the gas loss becomes important ($<15\%$ of the total gas mass). On the other hand, \cite{Steinhauser16} took a similar approach and studied different infalling orbits for three distinct clusters.
   They found that the SFR rises by up to 60\% for galaxies with V$_{max}$=170 km/s and $\log{M_{\ast}}\sim10.6$ in a cluster with similar properties to those we found in XMM2235-2557 after 0.5 Gyr and following an orbit that goes through the very central regions of the cluster. We translated the  SFR rise predicted by \citet{Ruggiero17} and \cite{Steinhauser16} into a change in B-band luminosity using the EzGal python code (\citealt{Mancone12}). EzGal is a tool that takes models of how the SED of a stellar population evolves with time and projects it through filters to calculate several physical properties, including magnitude evolution, as a function of redshift. In our case, we used the model libraries from \citealt{BC03} to study the evolution of the B-band luminosity evolution after a short starburst caused by the compression of the gas in the inner disk due to RPS. We find a brightening of 0.9 and 0.3 magnitudes for \citet{Ruggiero17} and \cite{Steinhauser16} conditions, respectively. 
   
    \begin{figure*}
      \centering
      \includegraphics[width=12cm]{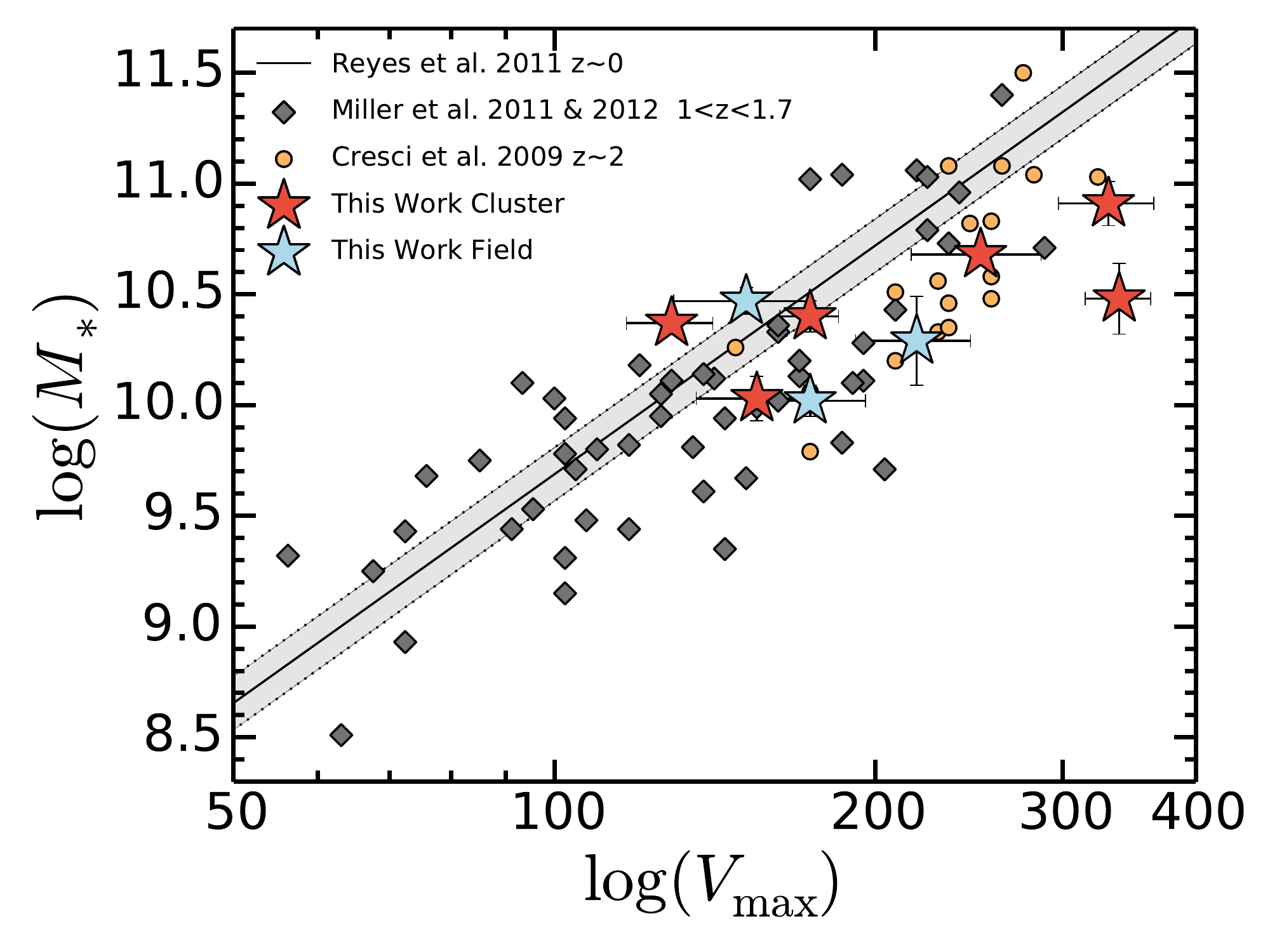}
      \caption{Stellar mass TFR. Red and blue stars are cluster and field galaxies in our sample, respectively. Gray diamonds show a sample of 42 field galaxies at 1$<$z$<$1.7 from \citet{Miller11} and \citet{Miller12}, who found a well-established M$_{*}$-TFR at this redshift. Orange circles are 18 field galaxies from \citealt{Cresci09} at z$\sim$2. The solid line is the local M$_{*}$-TFR from Fig. 23 in  \citet{Reyes11}, taking V$_{2.2}$ as V$_{max}$. The shaded area represents 1$\sigma$ deviation from the previous relation.}
      \label{SMTFR}
     \end{figure*}
   
   In summary, results from previous RPS studies and simulations point toward an enhancement of the SFR of the central regions of infalling cluster galaxies after going through the cluster environment for $<1Gyr$, but maintaining undisturbed velocity fields in the inner parts, as we find in our study. However, the variety of side effects playing a role in the process, such as inclination of the galaxy with respect to the infalling direction, density of the ICM, infalling velocity, and gas fraction, prevents us from extracting strong conclusions about the nature of the luminosity enhancement of our cluster galaxies without further observations. 
   
   \subsection{Stellar mass TFR}
   
   In virialized galaxies, a higher maximum rotation velocity can only be explained by a higher total mass value, including Baryonic (stellar plus gas) and dark matter content. Since it is not possible to obtain direct information about the gas fraction for distant galaxies, $M_{\ast}$ is the only available quantity we have to explore the evolution of the Baryonic mass in galaxies at this redshift. In Fig. \ref{SMTFR} we investigate the stellar mass TFR (M$_{*}$-TFR) for the field and cluster galaxies in our sample. Again, our cluster sample is composed of two groups divided by their total mass ($V_{max}$): three massive fast-rotating galaxies with $10.5>\log{M_{\ast}}>10.9$ embody the first group, while another three slow-rotating galaxies show slightly lower stellar masses, $10.0>\log{M_{\ast}}>10.4$. Our results follow the M$_{*}$-TFR for spiral galaxies established by \citealt{Miller12} at redshift 1.3<z<1.7  and are in agreement with previous studies at similar redshift, like \citealt{Miller11} z<1.3 and \citealt{Cresci09} at z$\sim$2. We compare our dataset with the local M$_{*}$-TFR established by \citealt{Reyes11} using a subsample of local galaxies from SDSS DR7.  
   
   While the B-band TFR is sensitive to recent episodes of star formation, the M$_{
   *}$-TFR tracks the overall evolution of the underlying stellar population. From combining the results from both incarnations of the TFR, we draw the following conclusions: high total mass cluster galaxies show stellar mass $\log{M_{\ast}}\geqslant10.5$ and lie within the 2$\sigma$ region of the local B-band TFR, while at the same time they are offset by $\langle\Delta\log{M_{*}}\rangle=-0.66$ with respect to the local $M_{\ast}$-TFR. On the other hand, the group of low-mass cluster galaxies show lower stellar mass, $\log{M_{\ast}}\leqslant10.5$, and are on average offset by 5 sigma toward higher B-band luminosity, with negligible deviations with respect to the local $M_{\ast}$-TFR, $\langle\Delta\log{M_{*}}\rangle=-0.05$.
   
   One possible explanation for this behavior would be that high-mass distant galaxies have grown their stellar mass following star formation histories that are compatible with quiescent evolution, excluding starbursting episodes in the last few billion years. In contrast, low-mass galaxies at this epoch are still in the early phases of assembling their disk and are more prone to suffer environmental effects that could enhance their SFR and thus, their B-band luminosity. The tendency for high-mass galaxies to develop their disk first has recently been studied by \citet{Simons16} for field galaxies at z$\sim$2 using MOSFIRE. The most massive galaxies in their sample ($\log{M_{\ast}}\geqslant10.2$) lie below the local $M_{\ast}$-TFR of \citet{Reyes11} and exhibit similar rotation support as their local counterparts ($V_{rot}/\sigma\geqslant2-8$), while at lower stellar masses, galaxies start to display small rotation support ($V_{rot}/\sigma\leqslant1$) and lie on the other side of the relation. In an environmental frame, the combination of the degree of rotational support ($V_{rot}/\sigma$) with tracers of current star formation might be a useful tool to interpret offsets in the different representations of the TFR.  
   

   \section{Conclusions} 
    
    Using the FORS2 instrument at the ESO Very Large Telescope, we have studied a sample of 25 galaxies in the XMMU2235-2557 field of view. We carried out a kinematic analysis for 6 cluster members at z$\sim$1.4 and 3 field galaxies at 1<z<1.2 and determined their maximum rotation velocity V$_{max}$. Structural parameters (such as disk inclination and scale length) were derived on HST/ACS and HAWK-I images. We analyzed the distant Tully-Fisher and velocity-size relations in XMM2235-2557 and compared them with reference samples at similar redshift and the local Universe, taking into account additional results from galaxy evolution simulations. Our main findings can be summarized as follows:

   \begin{enumerate}
      \item At given V$_{max}$, cluster galaxies are more luminous (in restframe B band) and smaller (in rest-frame z band) than their local counterparts toward higher redshifts. By z=1.4 we find for cluster members an average brightening of $\langle\Delta M_{B}\rangle=-1.6$ mag in absolute B-band magnitude and a decrease in size by a factor of $\sim$2-3.
      
      \item The cluster galaxies in XMM2235 were divided in two subsamples according to their V$_{max}$, occupying two different loci in the TFR. The first is composed of relatively slowly rotating (low total mass) galaxies that appear offset from the local TFR by $\sim 5\sigma$. The second lies within the 2$\sigma$ deviation region, in agreement with previous observational findings \citep{Boehm15} and semi-analytic models for field galaxies at similar redshift \citep{Dutton11}.
      The galaxies in our sample show smaller offsets and scatter in the stellar mass TF diagram than in the B-band TF diagram. The subsample of fast-rotating galaxies show $10.5>\log{M_{\ast}}>10.9,$ while the others have slightly lower stellar masses, $10.0>\log{M_{\ast}}>10.4$. Although both subsamples have a similar stellar mass, they might be affected differently by cluster-specific processes, which might enhance the SFRs and, in turn, B-band luminosities in the low-mass subsample.
      
      \item The origin of the TFR offsets for the group of slowly rotating (low total mass) galaxies is not clear. We have discussed several possibilities to explain our results, such as the underestimation of V$_{max}$ as a result of the shape of the RC, a magnitude bias in our sample, and a temporary brightening in the B-band luminosity of these galaxies caused by the interaction with the environment. A combination of the two latter options appears to be the most likely explanation for our findings. The effect of the Malmquist bias is limited and cannot be the sole explanation for the offsets of the slowly rotating subsample: $\langle\Delta M_{B}\rangle=-2.4$. Results from previous ram pressure stripping studies and simulations show that it is possible to enhance the SFR in the central regions of infalling cluster galaxies (and thus the B-band luminosity) by maintaining undisturbed velocity fields (and RCs) at smaller radii than 3-4 scale lengths, as we find in our sample. However, the small size of our cluster sample together with the variety of effects playing a role in this process, such as the inclination of the galaxy with respect to the infalling direction, density of the ICM, infalling velocity and gas fraction, and the combined possible effect of the magnitude bias prevent us from drawing firm conclusions about the nature of these offsets without further observations. 
      
      \item Analysis of the combined offsets in our sample from the Tully-Fisher and velocity-size relations reveal there is a correlation between them. Galaxies with a strong offset toward high B-band luminosity with respect to the local TFR have a similar size than their local counterparts at comparable V$_{max}$, while galaxies offset toward smaller sizes with respect to the local VSR have a B-band luminosity compatible with the local TFR. These results are in agreement with what was shown in a previous paper by \citealt{Boehm15}.
   \end{enumerate} 
   
In this paper we have explored the kinematics of galaxies in a high-redshift cluster. The distribution of our cluster subsamples in the TFR suggests that a population of galaxies exists that is consistent with the predicted evolutionary state of galaxies at this redshift, while cluster-specific interactions such as RPS might be responsible for the B-band luminosity enhancement suffered by the other half. However, it is not clear how likely this type of events is and whether the luminosity enhancement can be explained by a single process. Distinguishing between the origin of TFR offsets of bright cluster galaxies will require further work, such as examining the difference in star formation rate for distant cluster galaxies and the use of larger data sets covering a wide range of environments within the cluster.

   \begin{figure*}
     \centering
     \includegraphics[width=\textwidth]{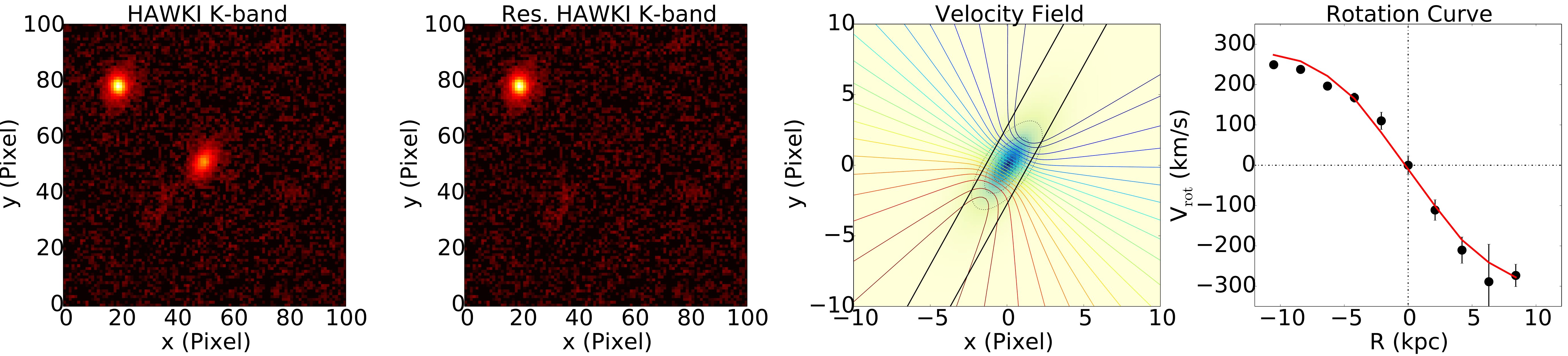}
     \includegraphics[width=\textwidth]{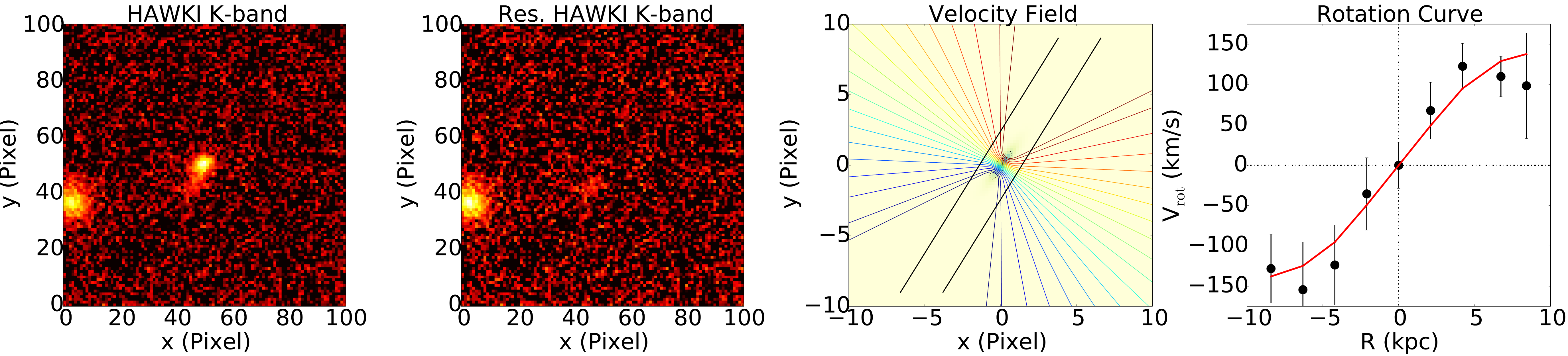}
     \includegraphics[width=\textwidth]{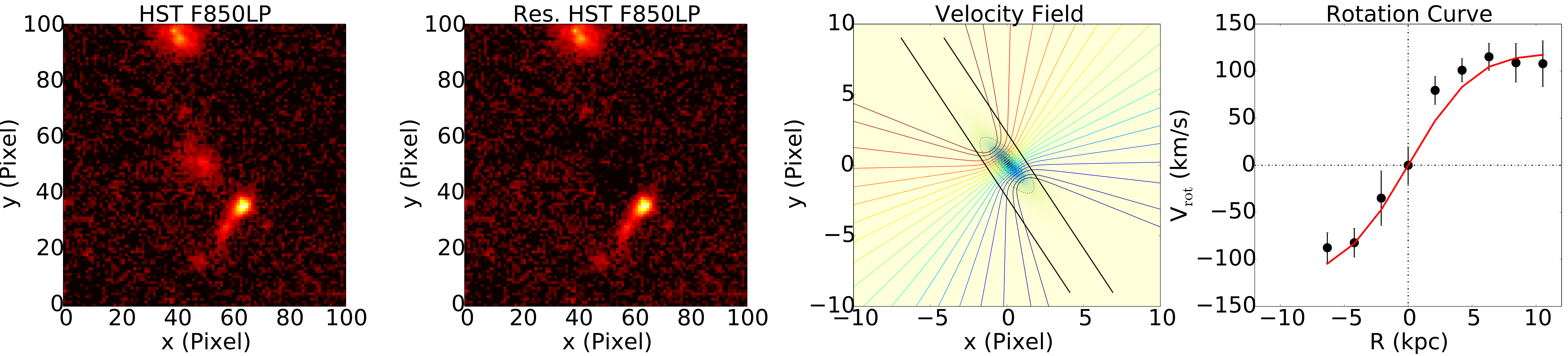}
     \includegraphics[width=\textwidth]{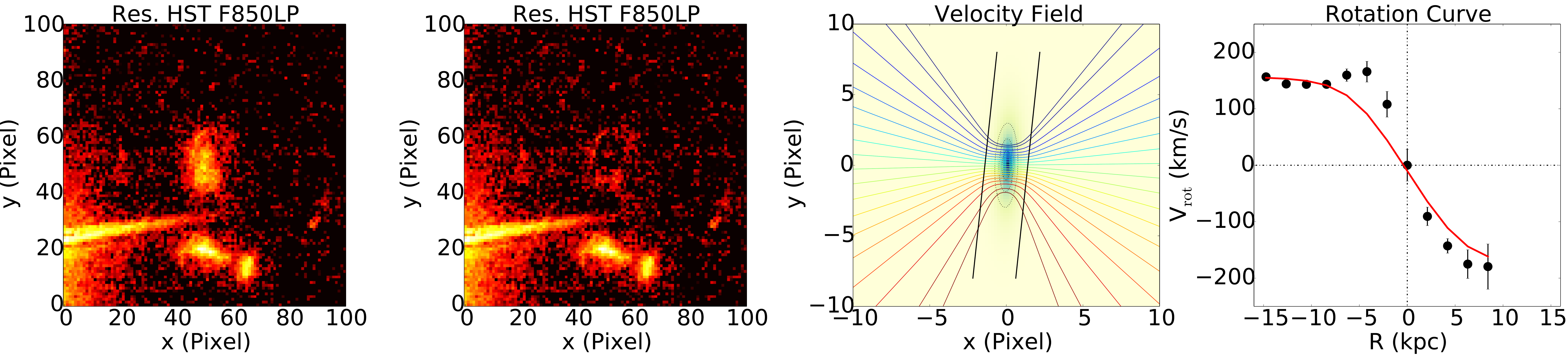}
     \includegraphics[width=\textwidth]{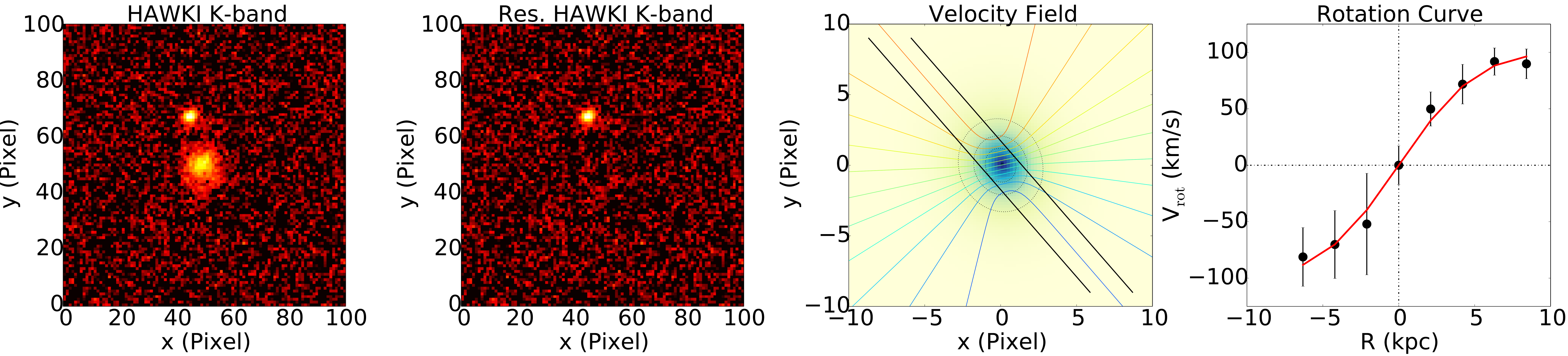}
     \caption[]{Our sample of cluster and field galaxies studied following the methods explained in Sect. \ref{SS:Methods} and presented in the same order as in Table \ref{tab7}. The first column shows the HAWKI K-band or HST-F850LP image centered on the target galaxy. The second column displays the residuals after subtracting the 2D model of the galaxy. The third column presents the synthetic velocity field based on the observed structural parameters after fitting the simulated rotation curve to the observed curve. The black lines mark the position of the edges of the slit. The fourth column displays the rotation curve (black points) in the observed frame, and the fitted simulated RC (red line).}
     \ContinuedFloat
     \label{foot}
   \end{figure*}

    \begin{figure*}
     \centering
     \includegraphics[width=\textwidth]{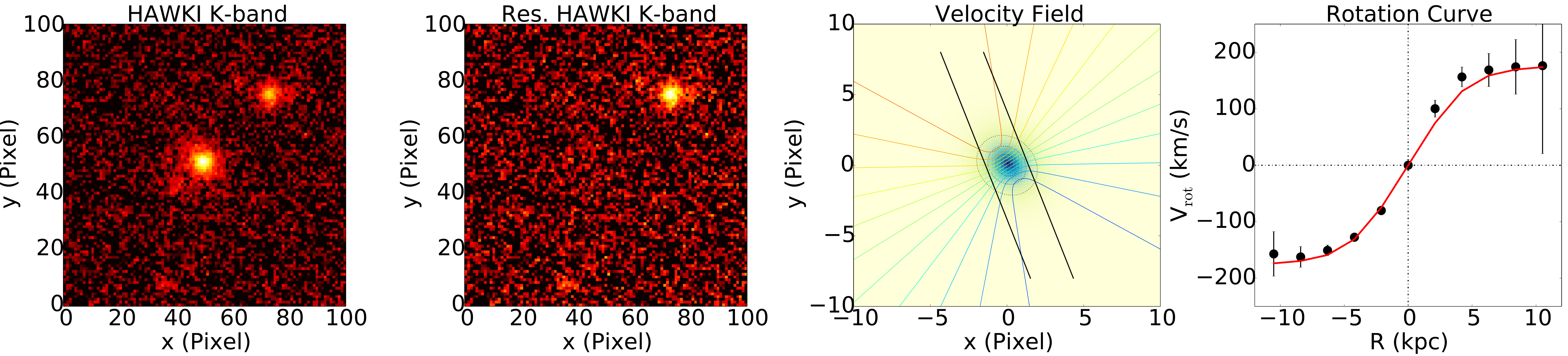}
     \includegraphics[width=\textwidth]{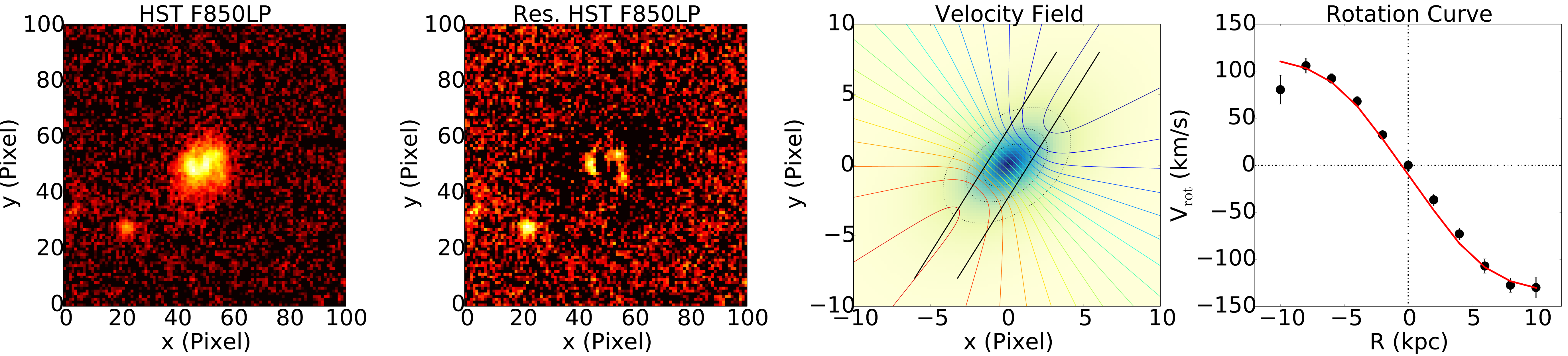}
     \includegraphics[width=\textwidth]{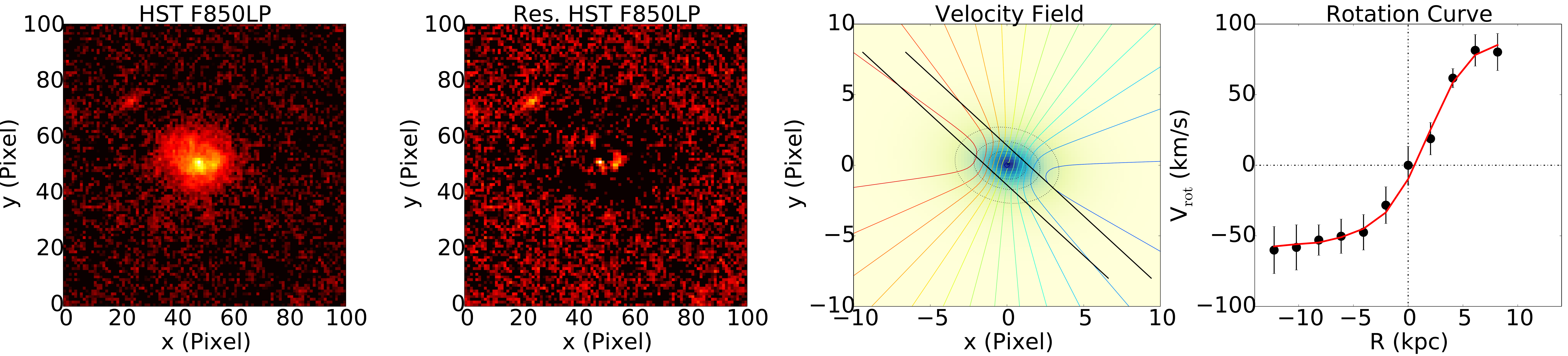}
     \includegraphics[width=\textwidth]{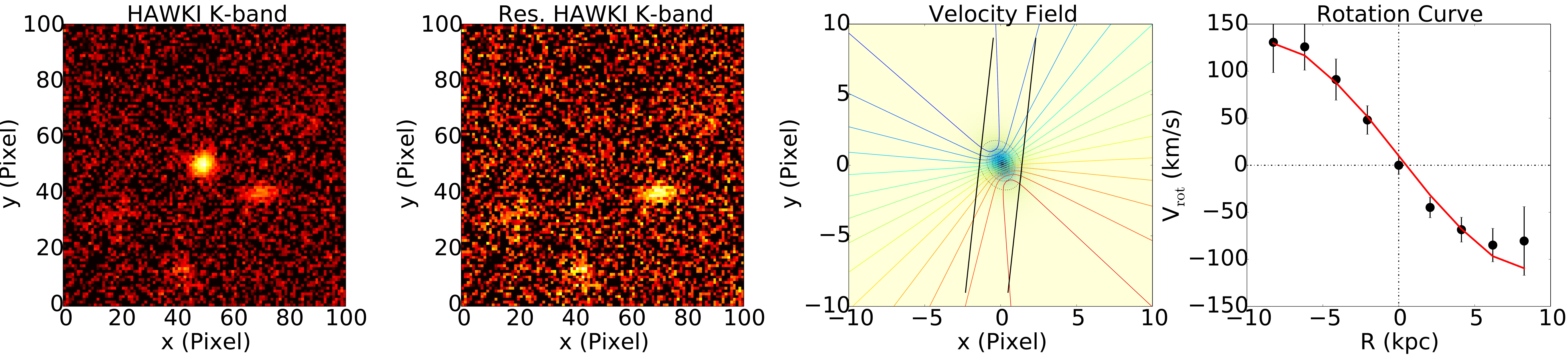}
     \caption[]{(Continued)}
    \end{figure*}

   \begin{table*}[h]
   \centering
\caption{General properties of our cluster and field samples. IDs, J2000 coordinates, redshift, AB absolute B-band magnitude before correcting for intrinsic dust absorption, observed K-band magnitude, $J-K$ color, and logarithmic stellar mass.}
\label{tabtot}
\begin{tabular}{cccccccccc}
\hline
\noalign{\vskip 0.1cm}
ID & RA & DEC & $z$ & $M_{B}$ & $m_{K}$ & $J-K$  & $\log{M_{\ast}/M_\odot}$ \\
   & (hh:mm:ss)  & (dd:mm:ss) &     & (mag) & (mag) & (mag) &  \\
\noalign{\vskip 0.1cm}
\hline 
\hline 
\noalign{\vskip 0.2cm}
1 & 22:35:21.6 & -25:54:30.4 & 1.364 & -22.91 $\pm$ 0.16 & 20.24 & 0.62 & 10.96 $\pm$ 0.12  \\ 
\noalign{\vskip 0.15cm}
2 & 22:35:21.7 & -25:54:39.4 & 1.391 & -22.50 $\pm$ 0.13 & 20.62 & 0.72 & 10.90 $\pm$ 0.08   \\ 
\noalign{\vskip 0.15cm}
3 & 22:35:33.5 & -25:55:08.7 & 1.366 & -22.87 $\pm$ 0.11 & 20.16 & 0.85 & 10.48 $\pm$ 0.16  \\
\noalign{\vskip 0.15cm}
4 & 22:35:21.9 & -25:55:38.9 & 1.391 & -22.46 $\pm$ 0.12 & 20.31 & 0.82 & 11.16 $\pm$ 0.06  \\
\noalign{\vskip 0.15cm}
5 & 22:35:33.1 & -25:55:47.1 & 1.358 & -21.82 $\pm$ 0.11 & 21.40 & 0.64 & 10.03 $\pm$ 0.10   \\
\noalign{\vskip 0.15cm}
6 & 22:35:27.1 & -25:56:34.7 & 1.386 & -21.78 $\pm$ 0.13 & 21.31 & 0.62 & 10.59 $\pm$ 0.05  \\
\noalign{\vskip 0.15cm}
7 & 22:35:19.4 & -25:56:56.0 & 1.380 & -21.64 $\pm$ 0.12 & 21.05 & 0.86 & 10.78 $\pm$ 0.06   \\
\noalign{\vskip 0.15cm}
8 & 22:35:21.8 & -25:57:14.0 & 1.399 & -20.96 $\pm$ 0.11 & 21.89 & 0.81 & 10.37 $\pm$ 0.05  \\
\noalign{\vskip 0.15cm}
9 & 22:35:21.6 & -25:57:38.6 & 1.389 & -20.33 $\pm$ 0.12 & 22.10 & 0.97 & 10.43 $\pm$ 0.05   \\
\noalign{\vskip 0.15cm}
10 & 22:35:33.0 & -25:57:57.4 & 1.390 & -22.36 $\pm$ 0.15 & 20.65 & 0.85 & 10.98 $\pm$ 0.14   \\
\noalign{\vskip 0.15cm}
11 & 22:35:18.1 & -25:58:06.6 & 1.382 & -22.11 $\pm$ 0.14 & 21.61 & 0.39 & 10.40 $\pm$ 0.07   \\
\noalign{\vskip 0.15cm}
12 & 22:35:17.1 & -25:58:35.4 & 1.395 & -22.14 $\pm$ 0.11 & 21.64 & 0.29 & 10.18 $\pm$ 0.11   \\
\noalign{\vskip 0.15cm}
13 & 22:35:26.0 & -25:58:53.5 & 1.388 & -22.57 $\pm$ 0.12 & 20.14 & 1.02 & 11.18 $\pm$ 0.06   \\
\noalign{\vskip 0.15cm}
14 & 22:35:21.0 & -25:59:03.8 & 1.395 & -21.56 $\pm$ 0.11 & 21.66 & 0.66 & 10.30 $\pm$ 0.09   \\
\noalign{\vskip 0.15cm}
15 & 22:35:27.8 & -25:59:48.8 & 1.353 & -22.17 $\pm$ 0.14 & 21.05 & 0.42 & 10.68 $\pm$ 0.03  \\
\noalign{\vskip 0.15cm}
16 & 22:35:17.1 & -26:00:26.3 & 1.357 & -22.60 $\pm$ 0.11  & 20.49 & 0.75 & 10.91 $\pm$ 0.10   \\
\noalign{\vskip 0.15cm}
17 & 22:35:21.1 & -26:01:22.8 & 1.397 & -21.33 $\pm$ 0.10 & 21.49 & 0.88 & 10.69 $\pm$ 0.13  \\
\noalign{\vskip 0.15cm}
\hline 
\noalign{\vskip 0.2cm}
18 & 22:35:23.0 & -25:54:58.8 & 0.560 & -19.26 $\pm$ 0.15 & 22.02 & 0.27 & 9.27 $\pm$ 0.11   \\ 
\noalign{\vskip 0.15cm}
19 & 22:35:24.8 & -25:55:27.0 & 0.994 & -21.38 $\pm$ 0.10 & 21.26 & 0.34 & 10.02 $\pm$ 0.07    \\
\noalign{\vskip 0.15cm}
20 & 22:35:17.8 & -25:56:05.2 & 0.765 & -23.00 $\pm$ 0.14 & 20.25 & 0.61 & 10.66 $\pm$ 0.05   \\
\noalign{\vskip 0.15cm}
21 & 22:35:22.2 & -25:56:20.0 & 1.511 & -21.83 $\pm$ 0.13 & 22.29 & 0.05 & 9.92 $\pm$ 0.05   \\
\noalign{\vskip 0.15cm}
22 & 22:35:21.5 & -25:57:30.2 & 1.091 & -21.47 $\pm$ 0.11 & 20.80 & 0.74 & 10.47 $\pm$ 0.06   \\
\noalign{\vskip 0.15cm}
23 & 22:35:17.7 & -25:59:14.3 & 0.884 & -20.64 $\pm$ 0.07  & 21.10 & 0.52 & 10.34 $\pm$ 0.12   \\
\noalign{\vskip 0.15cm}
24 & 22:35:28.8 & -26:00:10.3 & 0.986 & -21.36 $\pm$ 0.11 & 21.45 & 0.41 & 9.66 $\pm$ 0.06   \\
\noalign{\vskip 0.15cm}
25 & 22:35:19.4 & -26:01:08.1 & 1.167 & -21.33 $\pm$ 0.17 & 21.68 & 0.81 & 10.29 $\pm$ 0.20   \\
\noalign{\vskip 0.15cm}

\hline 
\end{tabular}
\break
\end{table*}

\begin{table*}[h]
\centering
\caption{IDs, redshift, structural parameters instrument ($Inst.$), intrinsic dust absorption ($A_{B}$), B-band-corrected luminosity ($M_{B_{corr}}$), and structural parameters of the cluster and field kinematic samples: scale length ($R_{d}$), inclination ($i$), position angle ($\theta$), misalignment ($\delta$), and logarithmic stellar mass.} 
\label{tab7}
\begin{tabular}{cccccccccccc}
\hline
\noalign{\vskip 0.1cm}
ID & $z$ &$Inst.$ & $A_{B}$ & $M_{B_{corr}}$ &$R_{d}$ & $i$ & $\theta$ & $\delta$ & $V_{max}$ & $\log{M_{\ast}/M_{\odot}}$  \\
     &  &  & (mag) & (mag)  & (kpc) & ($^{\circ}$) & ($^{\circ}$) & ($^{\circ}$) & (km/s)  &   \\
\noalign{\vskip 0.1cm}
\hline 
\hline 
\noalign{\vskip 0.2cm}
3 & 1.366 & HAWKI & -0.84 $\pm$ 0.12 & -23.71 $\pm$ 0.23 & 1.4 $\pm$ 0.3 & 65 $\pm$ 4 & -32 & 5  & 340 $\pm$ 24 & 10.48 $\pm$ 0.16 \\
\noalign{\vskip 0.15cm}
5 & 1.358 & HAWKI & -0.76 $\pm$ 0.20 & -22.58 $\pm$ 0.31 & 1.0 $\pm$ 0.2 & 75 $\pm$ 8 & -34 & 4 & 156 $\pm$ 19 & 10.03 $\pm$ 0.10\\
\noalign{\vskip 0.15cm}
8 & 1.399 & HST/ACS & -0.56 $\pm$ 0.15 & -21.52 $\pm$ 0.26 & 2.2 $\pm$ 0.4 & 70 $\pm$ 5 & 41 & 10 & 130 $\pm$ 12 & 10.37 $\pm$ 0.05\\
\noalign{\vskip 0.15cm}
11 & 1.382 & HST/ACS & -0.85 $\pm$ 0.17 & -22.96 $\pm$ 0.31 & 2.5 $\pm$ 0.5 & 76 $\pm$ 9 & 2  & 4 & 174 $\pm$ 11 & 10.40 $\pm$ 0.07 \\
\noalign{\vskip 0.15cm}
15 & 1.353 & HAWKI & -0.29 $\pm$ 0.07 & -22.46 $\pm$ 0.21 & 2.8 $\pm$ 0.6 & 44 $\pm$ 1 & 13 & 26 & 249 $\pm$ 35 & 10.68 $\pm$ 0.03\\
\noalign{\vskip 0.15cm}
16 & 1.357 & HAWKI & -0.19 $\pm$ 0.06 & -22.78 $\pm$ 0.18 & 1.9 $\pm$ 0.4 & 34 $\pm$ 1 & 34 & 14 & 334 $\pm$ 34 & 10.91 $\pm$ 0.10\\
\noalign{\vskip 0.15cm}
\hline 
\noalign{\vskip 0.2cm}
19 & 0.994 & HST/ACS & -0.30 $\pm$ 0.09 & -21.68 $\pm$ 0.19 & 1.5 $\pm$ 0.3 & 50 $\pm$ 2 & -46 & 16 & 172 $\pm$ 22 & 10.02 $\pm$ 0.07 \\
\noalign{\vskip 0.15cm}
22 & 1.091 & HST/ACS & -0.16 $\pm$ 0.06 & -21.64 $\pm$ 0.18 & 2.5 $\pm$ 0.5 & 40 $\pm$ 1 & 75 & 30 & 150 $\pm$ 22 & 10.47 $\pm$ 0.06 \\
\noalign{\vskip 0.15cm}
25 & 1.167 & HAWKI & -0.27 $\pm$ 0.07 & -21.60 $\pm$ 0.25 & 1.4 $\pm$ 0.3 & 44 $\pm$ 2 & -24 & 18 & 220 $\pm$ 36 & 10.29 $\pm$ 0.20 \\
\noalign{\vskip 0.15cm}
\hline 
\end{tabular}
\end{table*}

\begin{acknowledgements}

We acknowledge the thorough comments by the referee. This publication is supported by the Austrian Science Fund (FWF). A.B. is grateful to the Austrian Science Fund (FWF) for funding (grant number P23946-N16).

\end{acknowledgements}

\bibliographystyle{aa} 
\bibliography{references} 
\end{document}